\documentclass[nohyper,11pt,letterpaper]{JHEP3}
\usepackage{epsfig,yfonts}

\usepackage{amsfonts}
\usepackage{amsmath}

\newcommand{\beq}{\begin{equation}}
\newcommand{\eeq}{\end{equation}}
\newcommand{\beqa}{\begin{eqnarray}}
\newcommand{\eeqa}{\end{eqnarray}}
\newcommand{\beqar}{\begin{eqnarray*}}
\newcommand{\eeqar}{\end{eqnarray*}}

\def\be{\begin{equation}}
\def\ee{\end{equation}}
\def\ba{\begin{eqnarray}}
\def\ea{\end{eqnarray}}

\newcommand{\reef}[1]{(\ref{#1})}
\renewcommand{\eqref}[1]{(\ref{#1})}
\newcommand{\ssc}{\scriptscriptstyle}
\newcommand{\eg}{{\it e.g.,}\ }
\newcommand{\ie}{{\it i.e.,}\ }

\newcommand{\mt}[1]{\textrm{\tiny #1}}

\newcommand{\norm}[1]{\raise.3ex\hbox{:}#1\raise.3ex\hbox{:}}

\newcommand\prt{\partial}

\newcommand{\lsim}{\mathrel{\raisebox{-.6ex}{$\stackrel{\textstyle<}{\sim}$}}}

\newcommand{\p}{\phi}

\renewcommand{\t}{\theta}

\newcommand\cR{{\mathcal R}}
\newcommand\cK{{\mathcal K}}

\newcommand\MM{{\cal M}}
\newcommand\dM{{\partial\cal M}}
\newcommand\EE{{\cal E}}
\newcommand\OO{{\mathcal O}}

\newcommand\ls{\ell_s}  

\newcommand\gs{g_s} 
\newcommand\nf{N_5} 

\newcommand\x{\times}

\newcommand\dxbar{$\overline{\textrm{D8}}$}

\newcommand{\branes}{D8-$\overline{\textrm{D8} }$ pair~}

\def\det{{\rm det}}

\def\ap{\ell_s^2}

\newcommand\nts{\negthickspace}
\newcommand\bns{\nts \nts \nts}

\def\nc {N_\mt{c}}
\def\nf {N_\mt{f}}

\def\t6 {T_\mt{D6}}
\def\t8 {T_\mt{D8}}
\def\ut {u_\mt{KK}}
\def\utt{u_\mt{T}}
\def\u0{u_0}
\def\x4{{x^4}}
\def\tx4{{\tilde{x}{}^4}}
\def\hx4{{\hat{x}{}^4}}
\def\r4{{r_{\ssc \it 4}}}

\def\gym {g_\mt{YM}}
\newcommand{\nonsol}{D4-soliton}

\newcommand{\ct}{T_\mt{deconf}}
\newcommand{\uin}{{u_\infty}}
\newcommand{\Uin}{{U_\infty}}

\newcommand{\geff}{g_\mt{eff}}
\newcommand{\mq}{m_q}

\newcommand{\adss}[2]{\mbox{AdS$_{#1}\times {S}^{#2}$}}

\newcommand{\fc}{\frac}

\newcommand{\ra}{\rightarrow}
\newcommand{\sygn}{\omega}

\title{On quark masses in holographic QCD}

\author{Robert McNees$^a$, Robert C. Myers$^{a,b}$ and Aninda Sinha$^a$ \\
$^a$ {\it Perimeter Institute for Theoretical Physics, Waterloo,
Ontario N2L 2Y5, Canada}\\
$^b$ {\it Department of Physics and Astronomy, University of
Waterloo, Waterloo, Ontario}\\
\ \ {\it N2L 3G1, Canada}\\

\vskip .5cm

{\rm E-mail:}\ \ {\tt
rmcnees,$\,$rmyers,$\,$asinha@perimeterinstitute.ca}}

\abstract{Recently certain nonlocal operators were proposed to
provide quark masses for the holographic model of QCD developed by
Sakai and Sugimoto. The properties of these operators at strong
coupling are examined in detail using holographic techniques. We
find the renormalization procedure for these operators is modified
by the running of the five-dimensional gauge coupling. We explicitly
evaluate the chiral condensate characterized by these operators.}

\keywords{D-branes, Supersymmetry and Duality, Brane Dynamics in
Gauge Theories}

\preprint{arXiv:0807.nnnn [hep-th]}

\begin{document}{\vskip 1cm}

\section{Introduction}

Gauge/gravity dualities have proven to be a remarkable new framework
to study a large class of strongly coupled gauge theories
\cite{juan,bigRev}. However, the gauge theories that are currently
amenable to such holographic analysis are typically very different
from real world QCD. Hence constructing a holographic model of QCD
remains one of the most important challenges for this approach.
Currently, the most successful proposal is a construction by Sakai
and Sugimoto \cite{ss,ss2} based on a configuration of D8- and
$\overline{\textrm{D8}}$-branes in a D4-brane background. While
reliable calculations are limited to large $\nc$ and small
$\nf/\nc$, many observables seem to show a good approximation to
real QCD at low energies.

A key feature of the Sakai-Sugimoto model is that it exhibits the
desired non-Abelian chiral symmetry $U(\nf)_L\times U(\nf)_R$, as
well as its spontaneous breaking \cite{ss,ss2}. Of course, in real
world QCD, the analogous symmetry is only approximate as it is
explicitly broken by the quark masses. A shortcoming of the
D8/$\overline{\textrm{D8}}$/D4 model then is that the quarks are
precisely massless. While various suggestions have been made to
introduce quark masses \cite{mass2,tach1,tach2,tach3,tach4},
there remain technical difficulties in pursuing these proposals in
detail. A recent proposal which seems easier to study is based on
deforming the model with certain nonlocal operators
\cite{owl,mass1,old}. The underlying microscopic field theory is a
five-dimensional gauge theory where the chiral quarks are localized
on separate four-dimensional defects. Since the fermions of
different chiralities are separated in the five-dimensional
spacetime, no simple local mass term can be introduced in the UV
field theory. However, this spatial separation can be overcome by
connecting two quark fields with a Wilson line. Hence a natural
suggestion is to introduce a nonlocal operator to provide a quark
mass deformation \cite{owl,mass1,old}:
\beq  [\psi_L(x^\mu,\x4=0)]^a P \exp \left[i \int^{L/2}_{-L/2}\!
A_4\, d\x4 \right]^b_a [\psi_R(x^\mu,\x4=L)]_b\, .
 \label{oops0}
\eeq
As has been extensively studied for closed Wilson lines
\cite{maldaWilson,rey}, such a nonlocal operator would be dual to an
instantonic string worldsheet which extends between the
D8-$\overline{\textrm{D8} }$ pair. In the following, we examine the
properties of these operators in some detail.

An overview of the paper is as follows: in section
\ref{sec:Background}, we review the construction of the Sakai-Sugimoto background. In section \ref{wilson}, we consider the
nonlocal mass terms introduced in \cite{owl,mass1}. In particular,
we examine the affect of the dilaton coupling to the string
worldsheet. Even though this coupling only appears at higher order
in the $\alpha'$ expansion, we find that in the D4-brane background
it introduces a interesting modification in the renormalization of
the Wilson line. In section \ref{sec:SmoothWS}, we explicitly
calculate the expectation value of these nonlocal operators. In the
absence of any local fermion bilinears, this expectation value or
condensate is an order parameter characterizing the chiral symmetry
breaking in this holographic model. We close in section
\ref{discuss} with a discussion of our results and by making a few
observations about future directions. Appendix \ref{fluctuate}
provides the details of a calculation of the fluctuation determinant
of the worldsheet fields. The latter contributes at the same order
as the dilaton coupling but does not make any further modifications
of the renormalization of the nonlocal operators.

\section{Review of Sakai-Sugimoto background}
\label{sec:Background}

The Sakai-Sugimoto model \cite{ss,ss2} is based on the throat limit
of intersecting D4- and D8-branes, summarized by the array
\begin{equation}
\begin{array}{ccccccccccc}
   & 0 & 1 & 2 & 3 & 4& 5 & 6 & 7 & 8 & 9\\
\mbox{D4}: & \times & \times & \times & \times & \times & &  &  & & \\
\mbox{D8}: & \times & \times & \times & \times &  & \times & \times & \times & \times &\times   \\
\overline{\mbox{D8}}: & \times & \times & \times & \times &  & \times & \times & \times & \times &\times   \\
\end{array} ~.\label{D4D8}
\end{equation}
The world-volume theory of the $\nc$ D4-branes naturally gives rise
to a maximally supersymmetric $U(\nc)$ gauge theory in five
dimensions. Following \cite{thermalwitten}, the $x^4$ direction is
compactified and antiperiodic boundary conditions are imposed on the
fermionic fields around this circle. In the far infrared, one might
expect that the only relevant degrees of freedom arising from this
D4 world-volume theory correspond to four-dimensional Yang-Mills
with gauge group $SU(\nc)$. Further the intersection of the
D4-branes with $\nf$ D8-branes supports chiral fermions in the
fundamental representations of the gauge group $U(\nc)$ and of the
$U(\nf)$ flavour symmetry. These fermions propagate in the $3+1$
dimensions common to both sets of branes. Similarly, the
intersection with $\nf$ \dxbar-branes produces an analogous set of
four-dimensional anti-chiral fermions. Hence, the Sakai-Sugimoto
model produces a holographic description of QCD in the throat limit
of the D4-branes. The dual gravity theory in this framework yields
reliable results for large $\nc$ and strong 't Hooft coupling. Our
current understanding of this holographic model is limited to the
quenched approximation, \ie $\nf/\nc\ra0$, in which the D8-branes
are probes in the supergravity background.

\subsection{D4-brane background}\label{bak}

Here we review the supergravity background, which we refer to as the
the D4 soliton (following the nomenclature of \cite{soliton}). This
throat geometry for a stack of $N_{c}$ D4-branes with antiperiodic
fermions on the $\x4$ circle is the gravitational dual of a confined
phase of the $U(\nc)$ gauge theory \cite{thermalwitten}, as
described above. For comparison purposes, we also consider the
supersymmetric D4-brane throat with fermions that are periodic on
the $\x4$ circle. Both solutions can be expressed in the
form~{\footnote{The normalization for $F_4$ is different from what
is prevalent in the literature and has been chosen to be consistent
with the usual IIA action,
$I=\frac{1}{2\kappa^2}\int\,d^{10}x\sqrt{g}\left(e^{-2\phi}(R+4
(\nabla \phi)^2)-\frac{1}{48}F_4^2\right)\,.$}}
\begin{gather}
ds^{2}  =  \left(\frac{u}{R}\raisebox{14pt}{}\right)^{\frac{3}{2}}
\left(\raisebox{12pt}{} - dt^2 +\delta_{ij} \, dx^i dx^j + f(u)
(d\x4)^{2} \right) + \left( \frac{R}{u}\right)^{\frac{3}{2}} \left(
\frac{du^{2}}{f(u)} +
u^2 \, d\Omega_{\it 4}^{2} \right) \label{metric} \\
e^{\phi}  =  g_s \left( \frac{u}{R}\right)^{\frac{3}{4}} \qquad\qquad F_{\it
4} = 3\pi N_c \ell_s^3 \, \epsilon_{\it 4} \, . \label{metric1}
\end{gather}
The four noncompact directions of the gauge theory correspond to
$t=x^0$ and $x^i$ with $i=1,2,3$,  while the coordinate $\x4$ labels
the compact direction. The 56789-directions transverse to the
D4-branes are described by a radial coordinate $u$ and four angles
that parameterize a unit four-sphere. The $SO(5)$-invariant line
element on this sphere is $d\Omega_{\it 4}^2$, and the volume form
is $\epsilon_{\it 4}$. The function $f(u)$ is given by
 \be
 f(u) = 1-\frac{\displaystyle\ut^3}{\displaystyle u^3}
\, ,\label{solit}\\
 \ee
but the constant $\ut=0$ for the supersymmetric background.

The D4 soliton appears to have a conical singularity at $u=\ut$.
Regularity requires that the period of the compact direction, $\x4
\sim \x4 + 2\pi\, \r4$, is given by
 \be
2\pi\, \r4 = \fc{4 \pi}{3} \, \fc{R^{3/2}}{\ut^{1/2}}
\,. \label{deltatau}
 \ee
With this choice the $\x4$ circle smoothly shrinks to zero size at
$u=\ut$. Fermionic fields in the bulk must be antiperiodic on this
circle, reflecting the antiperiodic boundary condition on fermions
in the dual gauge theory. Unlike the soliton background, the
supersymmetric D4-brane geometry with $\ut=0$ exhibits a naked
curvature singularity at $u=0$. In that case there is no restriction
on the periodicity of the $\x4$ direction. Further, the dual gauge
theory is not confining.

The supergravity solution described above is completely specified
by the string coupling constant, $g_s$, the RR flux quantum (\ie the
number of D4-branes), $\nc$, and the non-extremality constant,
$\ut$. The remaining parameter is a length scale, $R$, which is
given in terms of these quantities and the string length, $\ell_s$,
by
 \be
R^{\,3} =  \pi g_s \nc\,\ell_s^{\,3}\,. \label{R}
 \ee
Various combinations of these parameters have direct interpretations
in the dual gauge theory. The holographic dictionary gives the
five-dimensional gauge coupling as $g_{\it{5}}^2 = (2\pi)^2 g_s
\ell_s$, so that the five-dimensional 't Hooft coupling is
 \be\label{tHooft5d}
  \lambda_{5} = g_{5}^{\,2}\,N_{c} ~.
 \ee
Since these couplings have dimensions of length there is a power-law
running of the dimensionless effective coupling \cite{many}
 \be
\geff^2=g_{\it{5}}^2\,\nc\,U\,,\label{run}
 \ee
where the energy scale $U$ is related to the radial coordinate $u$
in the D4-brane throat by $U =u/2\pi\ls^2$ \cite{Peet}. One finds
that the scale of Kaluza-Klein excitations of the compactified
coordinate $\x4$ gives the characteristic mass for glueballs
\cite{glue}
 \be
  M_\mt{KK} = \frac{1}{r_4} = \frac{3\,\ut^{\,1/2}}{2\,R^{3/2}} ~.
 \ee
Below this scale, the low-energy gauge coupling in four dimensions
is $\gym^2 = g_{\it{5}}^2 / 2\pi\,\r4$.

Supergravity provides a good description of physics in the D4-brane
background if two conditions are met. First, gravity calculations
are reliable if the length scale associated with spacetime
curvatures is small compared to the fundamental string tension. In
the D4 soliton solution \reef{solit} the Ricci scalar has a maximum
at $u \sim \ut$, where curvatures are of order $(\ut R^3)^{-1/2}$.
Hence we require
 \be
{\ut^{1/2} R^{3/2}\over\ell_s^2}  \gg 1 ~.
\label{hoof}
 \ee
In terms of gauge theory quantities this can be expressed as the condition
 \be
  \gym^2 \nc \gg 1
 \ee
so that the restriction to small curvatures corresponds to a large
't Hooft coupling in the effective four-dimensional gauge theory.
Second, string loop effects are suppressed as long as the local
string coupling is small: $e^\phi \ll 1$. The form of the dilaton
\reef{metric1} implies that, for finite values of the gauge theory
parameters, the inequality $e^{\phi} \ll 1$ can only be satisfied
over some finite range of the coordinate $u$. The string coupling
eventually becomes $\mathcal{O}(1)$ at a value of $u$ given by
 \be
  u_{crit} \simeq \fc{\nc^{1/3}\ell_s^2}{\gym^2\,\r4} \,. \label{crit}
 \ee
This critical radius naturally becomes large in the limit
$\gym\ra0$. Taken together, equations ~\reef{hoof} and \reef{crit}
indicate that the supergravity analysis in the \nonsol\ background
is reliable in precisely the strong-coupling regime of the 't Hooft
limit of the four-dimensional gauge theory.

In the strong coupling regime the QCD scale cannot be decoupled from
the compactification scale, \eg in the confining phase described by
the D4 soliton, the QCD string tension is $T \sim\gym^2 \nc/ \r4^2$
\cite{thermalwitten,holoQCD}. This means that, for most practical
purposes, calculations in the holographic framework are reliable in
a regime corresponding to a five-dimensional gauge theory. Since
this theory is nonrenormalizable it should be thought of as being
defined with a cut-off scale, $\Uin=\uin/2\pi\alpha'$. Above this
scale a UV completion with new degrees of freedom is required. This
completion may be a lift to M-theory, with the $x^{11}$ circle
opening up to reveal an asymptotically \adss{7}{4} background (with
identifications). On the field theory side of the duality, the UV
completion of the five-dimensional Yang-Mills theory is given by the
six-dimensional $(2,0)$ theory compactified on a circle. An
alternative UV completion would simply be type IIA superstring
theory in the asymptotically flat D4-brane background.

\subsection{D8-brane embeddings}

Our current understanding of the holographic model described in the
previous section is largely limited to the quenched approximation:
$\nf/\nc\ra0$. In this limit the D8-branes can be treated as probes
embedded in the supergravity background generated by the D4-branes
\footnote{See \cite{back}, for attempts to account for the
gravitational back-reaction of the D8-branes.}. The D8-brane fills
the noncompact 0123 directions as well as the angles on the $S^4$
transverse to the D4-branes. The nontrivial aspect of the embedding
is given by a function $\x4(u)$ that characterizes the D8-brane's
profile in the $u$-$\x4$ plane. With this choice of embedding, the
action for the D8-branes is
\beq I_\mt{D8} \sim -\nf\,\t8 \int du \, u^4 \sqrt{f(u)\,
(\prt_u{\x4})^2 + \left(\frac{R}{u} \right)^3\frac{1}{f(u)}} \,.
\label{D8action} \eeq
The resulting equation of motion for $\x4(u)$ is
\beq \frac{\prt}{\prt u} \left(\frac{u^4 f(u)\,
\prt_u{\x4}}{\sqrt{f(u)\, (\prt_u{\x4})^2
 + ({R}/{u})^3/f(u)}} \right)=0\, . \label{tauEom}
\eeq
The expression within the parentheses is constant. If we assume that
the profile is symmetric across the $u$-axis, crosses this axis at some value $u_0$, and is
smooth in the vicinity of this point, then this constant is given by
$\u0^4\sqrt{f(\u0)}$. The embedding equation can then be expressed
as
 \be
 \prt_u{\x4}=\fc{R^{3/2}\u0^4\sqrt{f(u_0)}}{u^{3/2}f(u)
 \sqrt{u^8f(u)-\u0^8\,f(u_0)}}\,. \label{ode}
 \ee
The boundary conditions for the D8-brane profile are: asymptotically
as $u\ra\infty$, $\x4\ra L/2$ and $\prt_u{\x4}\propto
1/u^{11/2}\ra0$; at the minimum $u\ra\u0$, $\x4\ra 0^+$ and
$\prt_u{\x4}\propto 1/(u-\u0)^{1/2}\ra\infty$. The full embedding
consists of two halves of this form. Hence the D8- and \dxbar-branes
are joined in a smooth profile at the minimum radius $\u0$ and the
two defects are separated by the asymptotic distance $L$ in the
$\x4$ direction. The limit $\u0\rightarrow\ut$ yields the trivial
embedding $\x4$=constant, which in the D4 soliton background
corresponds to a smooth joining of the D8- and \dxbar-branes with
asymptotic separation $L=\pi\,\r4$. In the supersymmetric background
($\ut = 0$) the trivial embedding is also allowed with an arbitrary
separation $L$. In this case, the D8- and \dxbar-branes terminate on
the singularity at $u=0$.

We can gain some intuition for these embeddings by considering the
supersymmetric background. In this case with $\ut=0$, $f(u)=1$ and
hence the embedding equation reduces to
\begin{equation}
\frac{\prt \x4}{\prt u} = \frac{R^{\,3/2}\,u_{0}^{\,4}}{u^{3/2}\,
\sqrt{u^{8} - u_{0}^{\,8}}}\,.\label{simpler}
\end{equation}
Integrating gives a solution in terms of an incomplete Beta function
\cite{njl}
\begin{equation}\label{SUSYEmbedding}
 \x4(u) = \frac{R^{\,3/2}\sqrt{\pi}}{8\,\sqrt{u_0}}\,\frac{\Gamma(9/16)}{\Gamma(17/16)} -
 \frac{R^{3/2}}{8\sqrt{u_0}}\,\beta\left(\frac{u_0^8}{u^8}, 9/16, 1/2\right) ~.
\end{equation}
This has a finite $u \to \infty$ limit, so that the asymptotic
(coordinate) separation of the $D8/\overline{D8}$ pair is given by
 \begin{equation}\label{asympsep}
L=\lim_{u \to \infty} 2\,\x4(u) =C_0\,\frac{R^{3/2}}{u_0^{\,1/2}}
\simeq 0.7245 \,\frac{R^{3/2}}{u_0^{\,1/2}}\,,
 \end{equation}
where $C_0\equiv\frac{\sqrt{\pi}\,\Gamma(9/16)}{4\,\Gamma(17/16)}$.
Of course, $L$ corresponds to the separation of the defects in the
dual gauge theory. With the D4 soliton background there is a maximum
separation corresponding to defects located at antipodes on the
$\x4$ circle. In the supersymmetric background the $\x4$ coordinate
need not be compact, so there is no restriction of this sort.
However, other considerations bound the maximum value of $L$ which
one might consider in this case. Notice that increasing $L$
corresponds to smaller values of $u_0$. If $u_0$ becomes too small,
the D8-brane extends into a region of high curvature and the
calculation described above is no longer reliable. Therefore one can
only reliably work with values of $L$ where the minimum of the brane
embedding is safely outside of this region.

With periodic fermions, the adjoint sector of the theory is
supersymmetric and the gauge theory is not confining. In the dual
background, free ``constituent'' quarks are realized as strings
stretching from $\u0$, the minimal radius of the branes, down to
$u=0$. The energy of these strings corresponds to the mass of the
constituent quarks: $\mq=\u0/2\pi\alpha'$. With \reef{asympsep},
this dynamically generated mass scale is given by
 \be
\mq =\fc{C_0^2}{8\pi^2}\,\fc{\lambda_5}{L^2}
=\fc{C_0^2}{4\pi}\,\gym^2\nc\fc{\r4}{L^2} \,. \label{mass}
 \ee
Since $2\pi\r4\ge L$ we always have $\mq\gg1/\r4$ in the strong
coupling regime, which reflects the fact that the infrared dynamics
does not decouple from the compactification scale. In the confining
background of the D4 soliton background, there are no free quarks
but one can still show that $\u0$ has the interpretation of roughly
determining the constituent quark mass $\mq$ as above \cite{show},
at least when $\u0$ is sufficiently larger than $\ut$.

The constituent quarks above are complicated bound states of
``current'' quarks (\ie the fundamental fields in the UV Lagrangian)
and adjoint fields. This is shown in a striking way by comparing the
quantum numbers of the constituent quarks to those of the current
quarks. In particular, the current quarks are singlets under the
global $SO(5)$ symmetry \cite{ss}. However, at $u_0$, the D8-brane
is wrapping the internal $S^4$ and the strings stretching from here
to $u=0$ in the supersymmetric background, can rotate in this
internal space. Hence the constituent quarks transform nontrivially
under $SO(5)$. Furthermore, it is likely that quantizing these
strings will give a spectrum of both bosonic and fermionic states.

One can also look at the low-lying meson spectrum by considering
excitations of the world-volume fields on the D8-branes
\cite{aharony,rowan}. One finds that this spectrum is (as expected)
characterized by the mass scale $\mq/\geff$ \cite{holom}, in
accord with the standard supergravity formula \cite{Peet}. Here, one
explicitly finds that mesons are both fermions and bosons \cite{ss},
rather than just bosons. The latter reflects the fact that these
infrared excitations are again complicated bound states of both the
(fermionic) quarks and (both fermionic and bosonic) adjoint fields
found in the UV Lagrangian.

\section{Nonlocal mass term} \label{wilson}

Recall that the underlying microscopic field theory is a
five-dimensional gauge theory with (chiral) fundamental matter
fields localized on two four-dimensional defects. These defects are
separated along the $\x4$ circle, so that fermions of different
chiralities live at different places in the spacetime. Hence a naive
mass term of the form
$[\psi_L(x^\mu,\x4=-L/2)]^a[\psi_R(x^\mu,\x4=L/2)]_a$ is not
possible --- in particular, it is not gauge-invariant. As described
in the introduction, the best one can do is to construct a nonlocal
but gauge-invariant operator \reef{oops0} with a Wilson line
connecting the quarks on the separated defects \cite{owl,mass1,old}.
This suggests that one consider the gravity/string dual as an
instantonic Euclidean string worldsheet which sits at $x^\mu$ and
extends between the D8-branes in the $\x4$ direction
\cite{maldaWilson,rey,drukker,fiol,loops}.\footnote{A similar class
of worldsheet instantons were studied with regard to the $U(1)_A$
problem in the Sakai-Sugimoto model \cite{oren}.} Of course, this
worldsheet does not quite reproduce the operator given above in
\reef{oops0}. Rather this holographic construction introduces an
``enhanced'' Wilson line which sources both the gauge field and the
adjoint scalars $\Phi^I$ of the five dimensional gauge theory
\cite{maldaWilson}
\beq \mathcal{O}(x^\mu) = [\psi_L(x^\mu,\x4=0)]^a P \exp \left[i
\int^{L/2}_{-L/2}\! \left(A_4-i\,n_I\,\Phi^I\right)\, d\x4
\right]^b_a [\psi_R(x^\mu,\x4=L)]_b\, .
 \label{oops}
\eeq
Here the (constant) normal vector $n_I$ indicates the position of
the worldsheet in the internal space. In principle, one could
consider an elaborate contour for the Wilson line connecting the two
fundamental fermions. However, for the sake of simplicity, we will
only consider a straight contour (with fixed $x^\mu$) in the
following.

In analogy with the usual holographic calculations of Wilson loops,
the expectation value of \eqref{oops} is given by
 \be \label{ExpectationValue}
\langle \mathcal{O} \rangle \sim e^{-I_\mt{WS}} ~,
 \ee
where $I_\mt{WS}$ is the worldsheet action for a string stretched
between the $D8$/$\overline{\textrm{D8}}$ pair. However, there is an
interesting difference between the present calculations and those
for the conformal super-Yang-Mills theory
\cite{maldaWilson,rey,drukker,fiol}. As described above, the
five-dimensional gauge theory under consideration here is defined
with a cut-off scale, $\Uin=\uin/2\pi\alpha'$. The necessity of this
cut-off is reflected in the nontrivial dilaton profile reaching
strong coupling at a large radius in the dual D4-brane background.
However, the standard Wilson line calculations are unaware of this
aspect of the physics since those calculations only consider the
leading-order Polyakov term for the worldsheet action in
\reef{ExpectationValue}. The effect of the nontrivial dilaton is
first seen at next-to-leading order in $\alpha'$, through the
Fradkin-Tsyetlin term \cite{Fradkin:1985ys}. Hence to better
understand the physics of Wilson line operators \reef{oops}, we are
motivated to carry the worldsheet calculations to first order in the
$\alpha'$ expansion. Consistency at this order demands that we also
include the fluctuation determinants of the worldsheet fields.

In the following, we find that the Fradkin-Tsyetlin contribution
generates new divergences in the calculation of
\eqref{ExpectationValue} and hence the renormalization procedure for
the worldsheet action must be modified. In particular, the new
divergences do \emph{not} seem to be removed by the Legendre
transform introduced in \cite{drukker}. We illustrate this in the
following sections by evaluating the action for a specific
worldsheet in the supersymmetric background. This simple calculation
exhibits the full set of divergences that appear in subsequent
calculations and so allows us to give a prescription for
renormalizing the worldsheet action to first order in $\alpha'$.

\subsection{Rectangular Worldsheet} \label{sec:RectWS}

To get a feeling for the issues that arise in the calculation of
\eqref{ExpectationValue}, we consider the trivial D8-brane
embedding, $\x4=\pm L/2$, in the supersymmetric D4-brane background.
This corresponds to having the D8- and $\overline{\rm D8}$-branes
extend straight down along the $u$ direction, from the cut-off at
$u_{\infty}$ to the curvature singularity at $u=0$. The gravity
approximation breaks down for small values of $u$, so we introduce
an (arbitrary) IR cut-off at $u=u_{\textrm{\sc ir}}$, with $R^{3/2}
u_{\textrm{\sc ir}}^{1/2} \gg \alpha'$. For convenience, we assume
that this is implemented by introducing a probe D4-brane where the
strings can end. The result is the rectangular worldsheet shown in
figure \ref{fig:1}. \FIGURE[ht]{
\includegraphics[width=.6\textwidth]{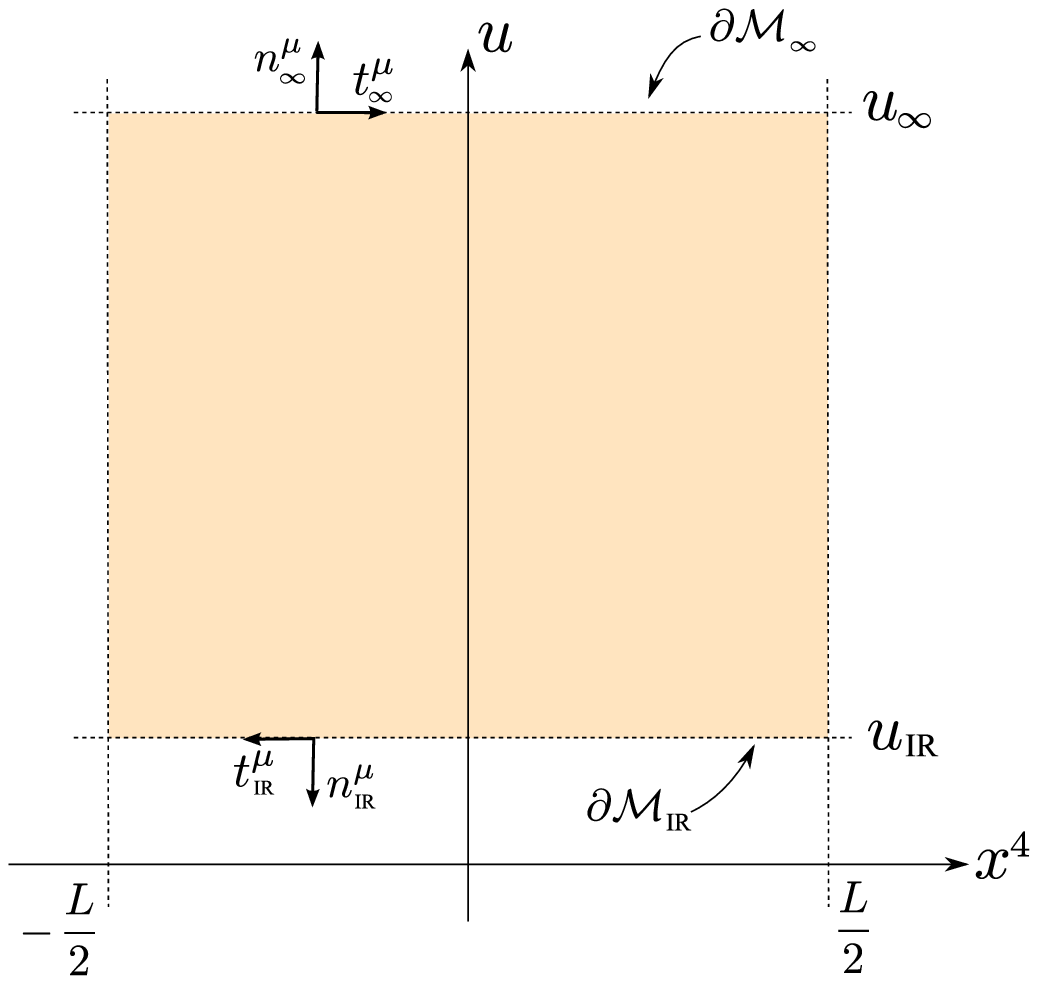}
\caption{The rectangular worldsheet described in the main text.}
\label{fig:1}}

Let us begin by considering the Polyakov action
 \be\label{actionx1}
I_\mt{P} = \frac{1}{4\pi\ap}\,\int_{\MM} \nts d^{\,2}\sigma
\sqrt{g}\,g^{ab}\,G_{IJ}\,\prt_aX^I \prt_bX^J\,,
 \ee
where $\MM$ is a worldsheet with boundary $\dM$, $g_{ab}$ is the
spacetime metric on $\MM$, and $G_{IJ}$ is the metric
\eqref{metric}. Two of the worldsheet scalars can be identified with
the coordinates $u$ and $\x4$ --- the remaining scalars can be
ignored for the moment. The worldsheet metric is taken to be the
same as the pullback of \eqref{metric} to $\MM$, which is given by
 \begin{equation}\label{WSmetric}
g_{ab} dx^{a} dx^{b} =
\left(\frac{u}{R}\right)^{3/2} (d\x4)^{2} +
\left(\frac{R}{u}\right)^{3/2} \nts du^{2} ~.
 \end{equation}
Evaluating the Polyakov action yields
 \be
I_\mt{P,rect} = \frac{L}{2\pi\,\ap}\,\left(\uin - u_{\textrm{\sc ir}} \right) ~.
 \label{rectaction1}
 \ee
The first contribution is a UV divergence proportional to the
cut-off scale $\uin$. This term is removed by the Legendre
transformation described in \cite{drukker}, which gives a
`renormalized' action
  \be
  I_\mt{P,rect}^\mt{(ren)} = -\frac{L}{2\pi\,\ap}\,u_{\textrm{\sc ir}} ~.
 \label{rnrectaction1}
 \ee
Alternately, subtracting the UV-divergent term from
\eqref{rectaction1} can be interpreted as a renormalization of the
field theory operator \eqref{oops}.

The Polyakov action is the leading order contribution to the
worldsheet action in the $\alpha'$ expansion. We must also take into
account terms at the next order in this expansion that couple to the
nontrivial dilaton of the D4-brane background. Specifically, we have
to evaluate the Fradkin-Tsyetlin term
\cite{Fradkin:1985ys,polchinski}
 \begin{eqnarray}
I_\mt{FT} & = & \, \frac{1}{4\pi}\,\int_{\MM} \!\! d^{\,2}\sigma
\sqrt{g}\,\cR(g) \, \Phi + \frac{1}{2\pi} \, \int_{\dM} \!\!\!\! ds
\,\cK\,\Phi
 + \frac{1}{2\pi}\,\sum_{i} \Phi(x^{\mu}_i)\, \left(\pi -
\theta_i \right)\,,  \label{actionx}
 \end{eqnarray}
where $\cR$ is the worldsheet Ricci scalar, $s$ is the proper
distance along the boundary $\dM$, and $\cK$ is the geodesic
curvature of the boundary. The latter is defined as
\beq \cK = -t^a n_b \nabla_a t^b \label{geodCurv} \eeq
where $t^{a}$ and $n^{a}$ are unit vectors tangent and normal to the
boundary, respectively. The last term in \eqref{actionx} is a sum
over corners where the embedding of the boundary is not smooth. A corner that
makes an angle $\theta$ gives a contribution proportional to $\pi -
\theta$, times the value of the dilaton at that point. These corner
terms can be thought of as arising from $\delta$-function
contributions to the geodesic curvature. Of course, with a constant
dilaton $\Phi_0$ the sum of all the terms in \reef{actionx} gives
$I_\mt{FT} = \chi\,\Phi_0$, where $\chi$ is the Euler character of
the worldsheet. The worldsheets that we consider all have the
topology of a disk, \ie $\chi=1$. Finally, consistency requires that
we take into account the fluctuation determinant on the worldsheet
at this order in $\alpha'$. This calculation is performed in
appendix \ref{fluctuate}, where we find that it does not make a
significant contribution to the worldsheet action. In particular,
these one-loop determinants do not generate any additional UV
divergences.

We now evaluate the individual terms in \reef{actionx}, beginning
with the scalar curvature term. The Ricci scalar for the worldsheet
metric \reef{WSmetric} is given by
\beq \cR(g)= -{3\over4}\,\frac{1}{R^{3/2}\,u^{1/2}} \label{wscurv}.
\eeq
and the first term in \reef{actionx} is
\beqar I_{\cR} &=&  \frac{1}{4\pi} \int_{-L/2}^{L/2} d\x4
\int_{u_\mt{IR}}^{\uin}du \left(-\frac{3}{4} \,\frac{1}{R^{3/2}\,u^{1/2}}
\log\left(\gs(u/R)^{3/4}\right) \right)  \nonumber   \\
&=& \frac{3\,L }{8\pi\,R } \left[\left(\fc{u}{R}\right)^{1/2}
\left(\frac{3}{2}-\log\left(\gs(u/R)^{3/4}\right)\right)
\right]^{\uin}_{u_\mt{IR}}. \label{expect} \eeqar
Next we consider the contributions to \reef{actionx} from the smooth
components of the boundary. The component of the boundary extending
from $(-L/2,\uin)$ to $(L/2,\uin)$ has tangent and normal vectors
given by
\beq t^a =(R/\uin)^{3/4}(\partial / \partial \x4)^a \, , \quad n^a =
(\uin/R)^{3/4} (\partial / \partial \uin)^a ~. \label{vecs}\eeq
Using these expressions in equation~\reef{geodCurv} gives the
geodesic curvature along this part of the boundary
\beq \cK = \frac{3}{4R}\left(\fc{R}{\uin}\right)^{1/4}~. \label{kk} \eeq
The proper distance along this edge is $ds = (\uin/R)^{3/4}d\x4$, so
the contribution to the action is
\beq I_{\cK} = \frac{1}{2\pi} \int_{-L/2}^{L/2}  ds\ \cK\, \Phi =
\frac{3\, L}{8 \pi\,R}\, \left(\fc{\uin}{R}\right)^{1/2}
\log\left[\gs \left(\frac{\uin}{R}\right)^{3/4} \right]\, .
\label{smooth} \eeq
The component of the boundary between $(-L/2,u_0)$ to $(L/2,u_0)$
makes a similar contribution; it differs by an overall minus sign
and the substitution $\uin\ra u_\mt{IR}$. The geodesic curvature
vanishes for the edges of the worldsheet along $\x4=\pm\,L/2$, so
they do not contribute to the action. Finally we consider the
contribution of the four corners of the worldsheet, each of which
makes an angle $\theta_i=\pi/2$. Their contribution to the action is
\beq I_\mt{corners} = \frac{1}{2} \, \log \left[
g_s\left(\frac{u_\mt{IR}}{R}\right)^{3/4} \right] + \frac{1}{2} \,
\log\left[ g_s \left(\frac{\uin}{R}\right)^{3/4}\right] ~.
\label{corners} \eeq
Collecting these terms, the action \reef{actionx} yields
\beq I_\mt{FT,rect} = \frac{9\,L}{16 \pi
R}\left(\sqrt{\fc{\uin}{R\,}} -
\sqrt{\fc{u_{\mt{IR}}}{R\,}}\,\right) + \frac{3}{8}\,\log
\left[\fc{\uin\,u_\mt{IR}}{R^2}\right] + \log \gs~. \label{fullS}
\eeq
Hence the inclusion of \eqref{actionx} leads to two new
UV-divergent terms in the worldsheet action, proportional to $\sqrt{\uin}$
and $\log \uin$. These are in addition to the
divergent term coming from the Polyakov action \reef{rectaction1}.
Notice as well that both \reef{expect} and
\reef{smooth} contained potentially divergent terms of the form
$\uin^{1/2}\log \uin$, however, these terms cancel out in the final
expression.

Although we have used a particularly simple background and
worldsheet configuration in the present calculation, the structure
of the UV divergences depends only on the asymptotic behaviour. This
means that the result obtained here is in fact universal, and the
divergences we have found also appear in more general situations.
The calculations in section \ref{sec:SmoothWS} --- both analytical
and numerical --- show this explicitly. Therefore, as we discuss
below, the results of this section lead to a general prescription
for renormalizing the worldsheet action and obtaining a finite
expectation value $\langle \mathcal{O} \rangle$.

As a final comment here, we note the term $\log\gs$ which arises in
\reef{fullS} from the inclusion of the Fradkin-Tseytlin term
\reef{actionx} in our calculation. Keeping the background scale $R$
fixed, \reef{R} gives $\gs\propto 1/\nc$ and hence one finds
$\langle{\cal O}\rangle\sim \nc$. Given that $\cal O$ is a bilinear
of fields in the fundamental representation of the gauge group, this
latter factor is precisely the expected result by the standard large
$\nc$ counting. Of course, this factor is a universal result for all
such worldsheet calculations, as we will see with the examples
calculated in the section \ref{sec:SmoothWS}.

\subsection{Renormalization of the Worldsheet Action} \label{sec:renorm}

One can try to address the UV divergences in \reef{fullS} by
applying the Legendre transform described in \cite{drukker}. The
authors there demonstrated that the `correct' action for observables
related to the minimal area of a string worldsheet is the Legendre
transform of \eqref{actionx1} with respect to some of the loop
variables --- see also \cite{fiol}. This is because some of the
worldsheet scalars satisfy Neumann boundary conditions
asymptotically rather than Dirichlet boundary conditions. Indeed, as
we commented above, implementing this Legendre transformation
removes the UV-divergent term from the Polyakov action
\eqref{rectaction1}. However, a straightforward application of the
same Legendre transform does not cancel the divergent terms in
\reef{fullS}.

To see that this is the case, first vary the full worldsheet action
with respect to the worldsheet fields. This gives an expression of
the form
 \be
\delta I =  \, \int_{\MM} \!\! d^{\,2}x \,\sqrt{g}\,\left[\,
\EE^{\mu\nu}\, \delta g_{\mu\nu} + \EE_{I} \,\delta X^{I}\raisebox{11pt}{}\right] +
\int_{\dM} \!\!\! dx \,\sqrt{h} \, \left[ \,\pi^{\mu\nu} \, \delta
h_{\mu\nu} + \pi_{I} \, \delta X^{I} \raisebox{11pt}{}\right]
 \ee
The coefficients of $\delta g_{\mu\nu}$ and $\delta X^{I}$ in the
integral over $\MM$ are the worldsheet equations of motion, while
the coefficients of $\delta h_{\mu\nu}$ and $\delta X^{I}$ in the
boundary integral are the momenta pulled back to $\dM$. The $\pi_I$
are given by
\begin{equation}\label{momentum}
\pi_{I} = \frac{1}{2\pi \ap} \, G_{IJ}\, n^{\mu} \partial_{\mu}
X^{J} + \frac{1}{2\pi} \, \cK\,\partial_{I} \Phi
\end{equation}
where $n^{\mu}$ is an outward pointing unit vector normal to $\dM$,
and all fields are evaluated at $\dM$. The Legendre transform of the
action with respect to some subset $\{X^{J}\}$ of the worldsheet
scalars is denoted $\widetilde{I}$ and is given by
 \ba\label{LT}
\widetilde{I} = I - \int_{\dM} \bns dx \,\sqrt{h} \,\sum_{\{X^{J}\}}
\pi_{J} \, X^{J} ~.
 \ea
Following \cite{drukker}, we construct the Legendre transform
of $I_{rect}$ with respect to the worldsheet scalar $u$ at
$u_{\infty}$
\begin{equation}
\widetilde{I}_{rect} = I_{rect} - \int_{\dM_{\infty}} \bns \nts \nts
dx \,\sqrt{h}\,\pi_{u}\,u ~.
\end{equation}
Using \eqref{momentum}, we have
 \ba
\pi_{u} = \frac{1}{2\pi\ap} \, \left(\frac{R}{u}\right)^{3/4} +
\frac{9}{32\,\pi}\,\frac{1}{R^{\,3/4} u^{5/4}}
 \ea
The induced metric at $u_{\infty}$ yields $\sqrt{h} =
(\uin/R)^{3/4}$ and so
\begin{equation}
\int_{\dM_{\infty}} \bns dx \,\sqrt{h}\,\pi_{u}\,u =
\frac{L}{2\pi\,\ap}\,u_{\infty} +
\frac{9\,L}{32\,\pi\,R^{\,3/4}}\,\sqrt{u_{\infty}} ~.
\end{equation}
The first term cancels the leading power-law divergence coming from
the Polyakov term \eqref{rectaction1}, but the second term does not
cancel the corresponding $\uin^{1/2}$ term in \reef{fullS}
 \ba\label{LTrectaction}
 \widetilde{I}_{rect} = -\frac{L}{2\pi\,\ap}\,u_{0} +
 \frac{9\,L}{32\,\pi\,R^{\,3/2}}\,(\sqrt{u_{\infty}} -
 2\,\sqrt{u_0}) +
 \frac{1}{2}\,\left(\Phi_{\infty} + \Phi_{0} \right) ~.
 \ea
Thus the usual Legendre transform of the worldsheet action does not
address the divergent terms at next-to-leading order in $\alpha'$.
As the structure of the UV divergences is universal, this approach
also fails for more general curved embeddings, such as those that we
study in the next section.

The simplest method for dealing with the divergences is to subtract
the terms in \reef{fullS} that depend on $\uin$. This approach is
closer in spirit to that applied in the holographic renormalization
of probe D-brane calculations, \eg see \cite{karch,brain}. We have
already noted that subtracting the $\uin$ term in the Polyakov
action can be interpreted in the field theory as a UV
renormalization of $\mathcal{O}$ and the same interpretation applies
to the new terms at next-to-leading order in the worldsheet action.
Such a subtraction is straightforward for the $\uin^{1/2}$ term,
however, we also have to deal with the logarithmic contribution from
the two corners at $u=\uin$, $\x4 = \pm L/2$. An ambiguity naturally
arises here because the subtraction, which takes the form
$3/4\,\log(\uin/u_\mt{sub})$, requires the introduction of a
subtraction scale $u_\mt{sub}$. Thus, our proposal for renormalizing
the worldsheet action is
 \be
 I_\mt{WS}^\mt{(ren)} = I_\mt{WS} - \frac{L}{2\pi\,\ap}\,u_{\infty} -
 \frac{9\,L}{16 \pi R}\,\sqrt{\frac{\uin}{R}} -\frac{3}{8}\, \log
 \left(\fc{\uin}{u_\mt{sub}}\right) \,.
 \label{fullir}
 \ee
The UV divergences that we subtract are universal and render the
worldsheet action finite up to terms of order $\alpha'$. As shown in
appendix \ref{fluctuate}, there are no divergences associated with
the fluctuation determinant. In the next section we apply this
renormalization to the worldsheet action for a string stretching
between the $D8$- and $\overline{D8}$-branes with the curved embedding.

\section{Worldsheet for smooth \branes embedding}
\label{sec:SmoothWS}

Now we turn to an explicit calculation of the expectation value
$\langle\mathcal{O}\rangle$ for the Sakai-Sugimoto model
\cite{ss,ss2}. As this holographic model does not permit the
construction of a local fermion bilinear, this expectation value is
the condensate which characterizes the spontaneous breaking of
chiral symmetry in this model. As described in section
\ref{sec:Background}, this spontaneous symmetry breaking is realized
in the gravitational dual by the \branes joining together to form a
smooth U-shaped embedding, as illustrated in figure \ref{fig:2}. In
the figure, the light blue region is the (Euclidean) worldsheet of a
string stretching between the $D8$-$\overline{D8}$ pair. The
boundary of this worldsheet consists of two smooth components: the
segment $\dM_{\infty}$ defined by the cut-off $u_{\infty}$, and the
segment $\dM_{D8}$ defined by the embedding $x^4 = x^4(u)$.

\FIGURE[ht]{
\includegraphics{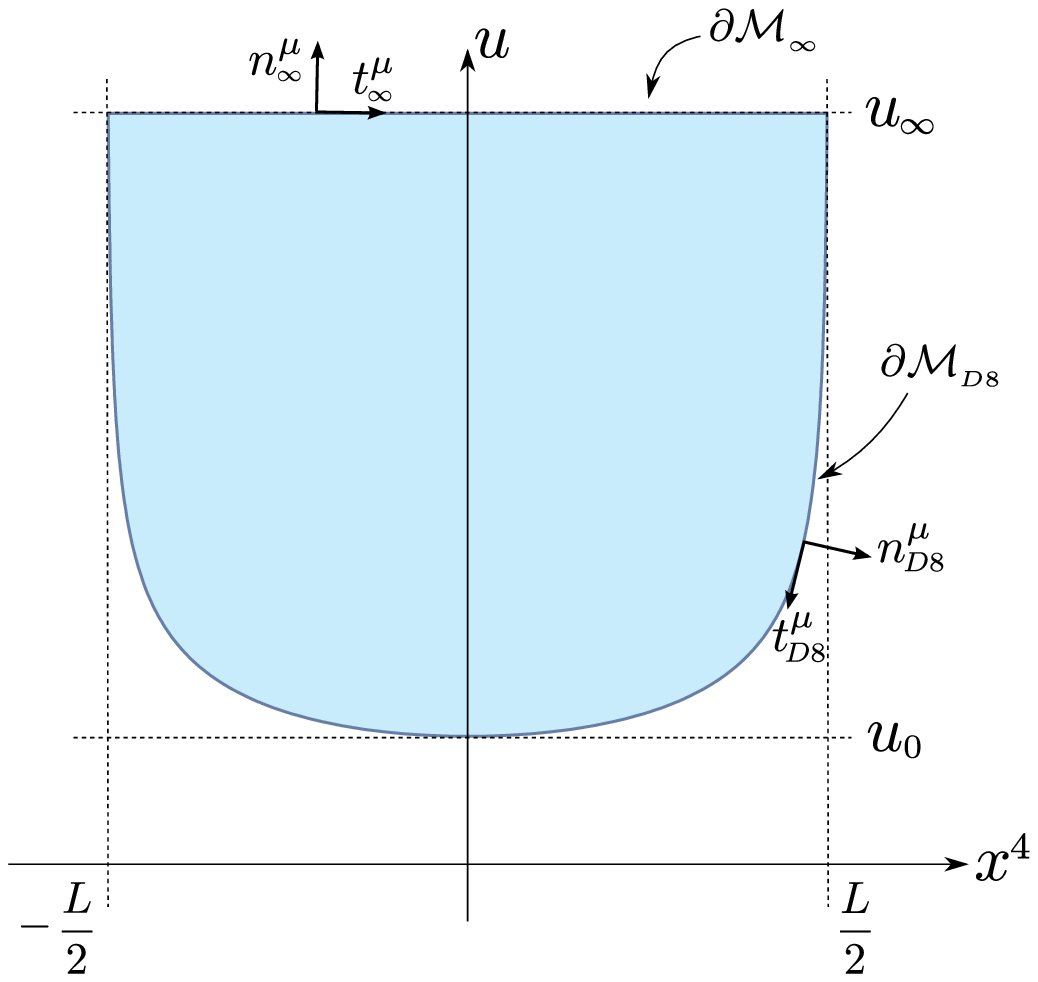}
\caption{Worldsheet for a smooth embedding of the $D8/\overline{D8}$
branes.} \label{fig:2}}
%


Recall that embedding equation \reef{ode} determines the D8-brane
profile $x^4(u)$
\begin{equation}\label{EmbeddingEqn}
\frac{\partial x^4}{\partial u} =
\frac{R^{3/2}\,u_{0}^{\,4}\,\sqrt{f(u_0)}}{u^{3/2}\,f(u)}\,\frac{1}{\sqrt{u^{8}\,f(u)
- u_{0}^{\,8}\,f(u_0)}}  ~.
\end{equation}
where $u_0$ is the minimum radius where the \branes joins smoothly
--- see figure \ref{fig:2} --- and $f(u) = 1 -
\left(u_\mt{KK}/u\right)^3$. With $\ut\ne0$, this equation cannot be
solved analytically and so the embedding must be determined
numerically.

To numerically solve for the embedding we define the following
dimensionless variables
\begin{equation} \label{DimensionlessVariables}
  z = \frac{u}{u_{0}} \quad \quad \sygn = \frac{u_\mt{KK}}{u_{0}} \quad \quad
  \psi = \sqrt{\frac{u_{0}}{R}}\, \frac{\,x^4}{R} ~,
\end{equation}
with $u_{0}$ the point on the $u$-axis where the embedding reaches
its minimum value. The restriction $u_{0} \leq u \leq u_{\infty}$
implies $1 \leq z \leq z_{\infty}$. Similarly, the parameter $\sygn$
takes values $\sygn \in [0,1]$. Here the lower limit corresponds to
$u_0\ra\infty$ but this limit is also realized in the extremal
background with $u_\mt{KK} =0$. The upper bound is reached when the
embedding reaches the minimum radius of the background at
$u=u_\mt{KK}$. In terms of these dimensionless variables, the
embedding equation becomes
\begin{equation}\label{DimensionlessEqn}
\frac{\partial \psi(z,\sygn)}{\partial z} =
\frac{z^{3/2}\,\sqrt{1-\sygn^{3}}}{z^{3}-\sygn^{3}} \,
\frac{1}{\sqrt{z^{8}-\sygn^{3}\,z^{5} - 1 + \sygn^{3}}} ~.
\end{equation}
Solving this equation numerically yields a family of embeddings
$\psi(z,\sygn)$ parameterized by $\sygn$.

With the standard dictionary $U =u/2\pi\ls^2$, the parameter $\sygn$
becomes a ratio of scales $\sygn = \frac{U_\mt{KK}}{U_{0}}$ where
$U_\mt{KK}$ and $U_0$ can be thought of as the confinement and
chiral symmetry breaking scales, respectively. In this model, these
are both dynamically generated scales determined by the fundamental
gauge theory parameters. For example, using \reef{deltatau} and the
subsequent formulae in section \ref{bak}, $U_\mt{KK}=\frac{2}{9}\,
\fc{\lambda_5}{(2\pi\,\r4)^2}$ and in the supersymmetric background,
$U_0=\mq$ in \reef{mass}. Hence, in principle $\sygn$ is also a
function of the parameters $L$, $\r4$ and $\lambda_5$. In fact, a
relatively simple expression can be derived by first noting that for
a sufficiently large cut-off $z_{\infty}=u_{\infty}/u_0$,
$\psi(z_\infty,\sygn)$ is essentially only a function of $\sygn$.
Then the asymptotic separation $L = 2\,x^{4}(u_{\infty})$ can be
expressed in terms of $\psi_{\infty}(\sygn)$ with
\begin{equation}
  L = 2\,\psi_{\infty}(\sygn)\,\sqrt{\frac{R^3}{u_0}} ~.
  \label{noname}
\end{equation}
Using the various expressions in section \ref{bak}, we then find
\begin{equation}\label{L_vs_uKK_relation}
\frac{1}{3}\,\frac{L}{\r4} =  \sqrt{\sygn}\,\psi_{\infty}(\sygn)
\end{equation}
and so in fact $\sygn$ is independent of the coupling $\lambda_5$.
Hence the coupling dependence of the individual scales $U_\mt{KK}$
and $U_0$ has canceled in the ratio defining $\sygn$. The function
$\sqrt{\sygn} \, \psi_{\infty}(\sygn)$ is shown in figure
\ref{fig:ImportantRelation} on the range $\sygn \in [0,1]$.
\FIGURE[ht]{
\includegraphics[width=.6\textwidth]{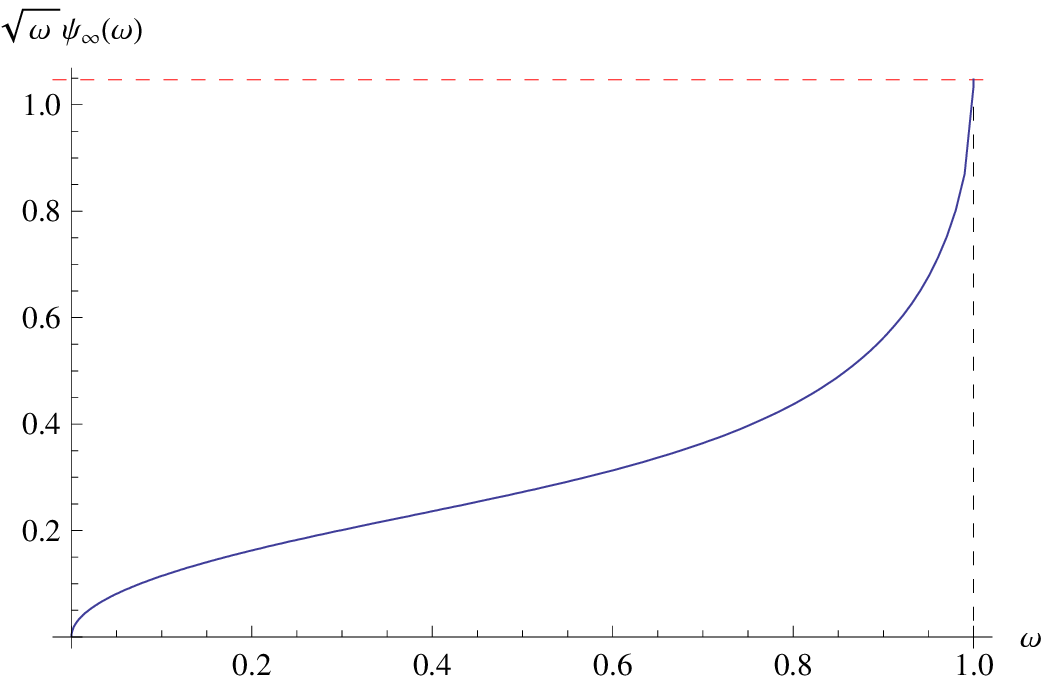}
\caption{The function $\sqrt{\sygn}\,\psi_{\infty}(\sygn)$ that appears on the right-hand-side of equation
\eqref{L_vs_uKK_relation}. The dashed red line corresponds to
$L = \pi\,r_4$, where the defects sit at antipodal points on the $\x4$ circle. The
dashed black line at $\sygn = 1$ is the bound $u = u_\mt{KK}$.}
\label{fig:ImportantRelation}}

Holding $L$ fixed in \eqref{L_vs_uKK_relation} implicitly gives
$\sygn$ as a function of $\r4$. This allows us to interpret
$\psi(z,\sygn)$ as a family of $D8/\overline{D8}$ embeddings with
constant asymptotic separation in $D4$-brane backgrounds with
different compactification radii $\r4$. Three such embeddings are
shown in figure \ref{fig:Family}. Alternately, fixing $\r4$ in
\eqref{L_vs_uKK_relation} implicitly gives $\sygn$ as a function of
$L$. In that case $\psi(z,\sygn)$ represents a family of
$D8/\overline{D8}$ embeddings with varying asymptotic separation in
a fixed, non-extremal $D4$-brane background.
\FIGURE[ht]{
\includegraphics[width=.6\textwidth]{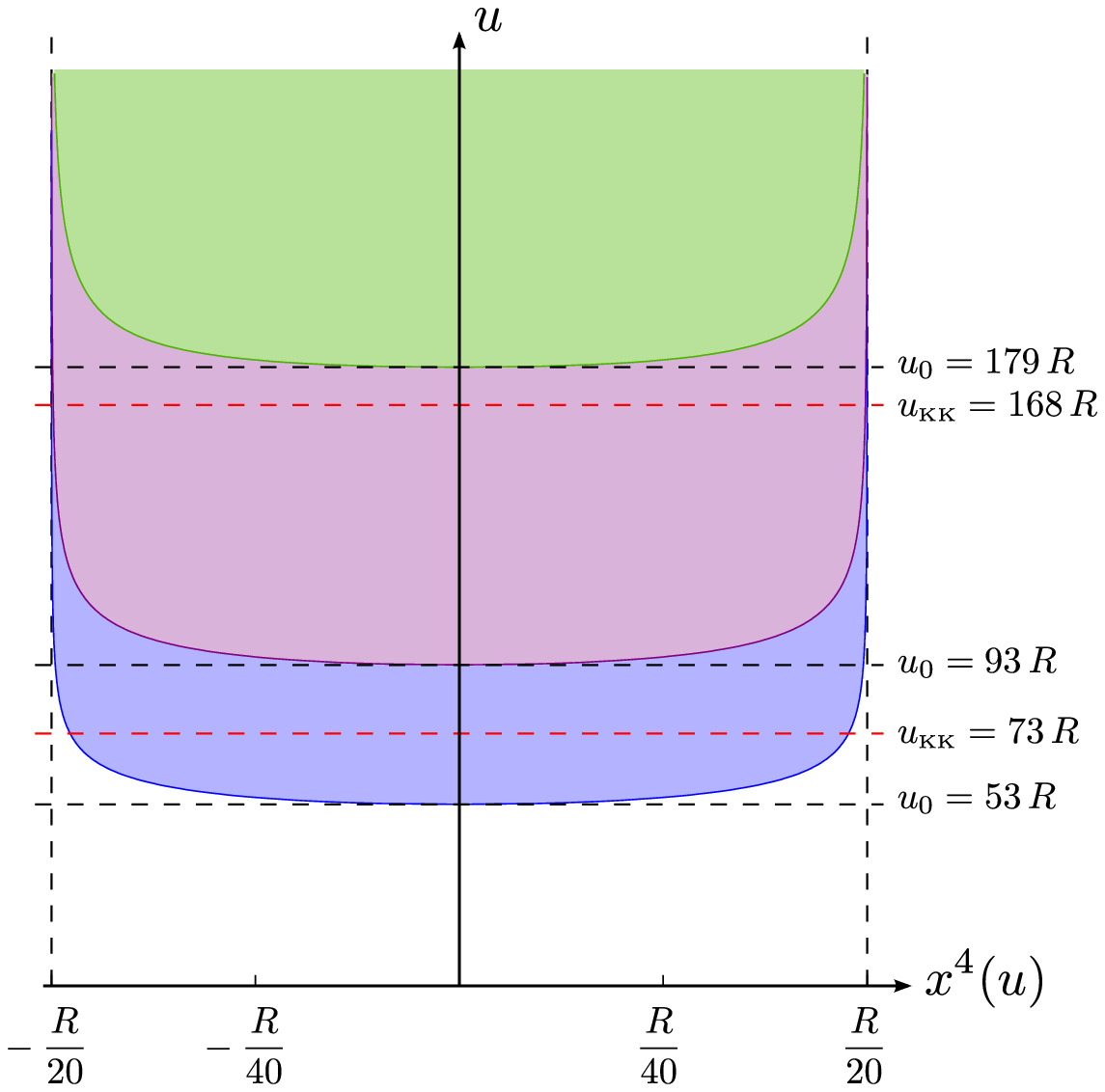}
\caption{$D8/\overline{D8}$ embeddings with fixed asymptotic
separation $L = R/10$, for $u_\mt{KK}=0$ (blue), $u_\mt{KK}=73\,R$
(violet), and $u_\mt{KK}=168\,R$ (green). The non-zero values of
$u_\mt{KK}$ are indicated by red dashed lines, and the minimum value
of $u$ reached by each embedding is indicated by a black dashed
line.} \label{fig:Family}}

\subsection{Worldsheet Action for the Curved $D8$-$\overline{D8}$ Embedding}
\label{pogo}

Next we explicitly evaluate the worldsheet action for a string
stretched between $D8$- and $\overline{D8}$-branes for the curved
embeddings described in the previous section. In the extremal
background the calculation can be performed analytically; in the
non-extremal case we must use numerical methods. While its
contributions are subdominant, we include the Fradkin-Tseytlin term
in the following for illustrative purposes.

As in the previous calculation, we identify the coordinates of the
Euclidean worldsheet with the spacetime coordinates $u$ and $x^{4}$.
The worldsheet metric and dilaton are given by
\begin{equation}\label{WSmet_and_dil}
  ds^{2} = \left(\frac{u}{R}\right)^{3/2} \nts f(u)\,(dx^4)^{2} +
   \left(\frac{R}{u}\right)^{3/2} \nts \frac{du^{2}}{f(u)}
  \quad \quad  \quad \Phi = \log \left( g_s\, \left(\frac{u}{R}\right)^{3/4} \right)\,.
\end{equation}
To compute the Fradkin-Tseytlin part of the action we need
expressions for the scalar curvature on $\MM$, as well as the proper
distance and geodesic curvature on $\dM$. The Ricci scalar for the
metric \eqref{WSmet_and_dil} is
\begin{equation}
 \cR = - \frac{3}{4\,R^{3/2} \, u^{1/2}} +
 \frac{15\,u_\mt{KK}^{\,3}}{4\,R^{3/2}\,u^{7/2}}~.
\end{equation}
The normal vector, tangent vector, and geodesic curvature for the
component of the boundary at $u = u_{\infty}$ are given by
\begin{equation}
n_{\mu} = \left(\frac{\uin}{R}\right)^{3/4} \nts
\frac{1}{\sqrt{f(\uin)}} \, \delta_{\mu}^{\,\,\, u}
\quad \quad \quad t_{\mu} = \left(\frac{R}{\uin}
\right)^{3/4}\nts \sqrt{f(\uin)} \, \delta_{\mu}^{\,\,\,x^{4}}
\end{equation}
\begin{equation}
{\cal K} =
\frac{3\,(\uin^{3}+u_\mt{KK}^{\,3})}{4\,R^{3/4}\,\uin^{\bns 13/4}\,\sqrt{f(\uin)}} ~.
\end{equation}
On the component of $\dM$ described by the embedding $x^{4}(u)$
these quantities are
\begin{equation}
n_{x^4} = \pm \left(\frac{u}{R}\right)^{3/4}
\frac{\sqrt{u^{8}\,f(u) - u_{0}^{\,8}\,f(u_0)}}{u^4} \quad \quad
n_{u} = - \left(\frac{R}{u}\right)^{3/4}
\left(\frac{u_0}{u}\right)^{4} \frac{\sqrt{f(u_0)}}{f(u)}
\label{bin1}
\end{equation}
\begin{equation}
 t_{x^4} = -\left(\frac{u}{R}\right)^{3/4}\left(
 \frac{u_0}{u}\right)^{4} \sqrt{f(u_0)}
 \quad \quad
t_{u} = \mp \left(\frac{R}{u}\right)^{3/4} \frac{\sqrt{u^{8}\,f(u) -
u_{0}^{\,8}\,f(u_0)}}{u^4 \, f(u)} \label{bin2}
\end{equation}
\begin{equation}
 {\cal K} = \frac{13\,u_{0}^{\,4}\,\sqrt{f(u_0)}}{4\,R^{3/4}\,u^{17/4}} ~.
\end{equation}
Above in \reef{bin1} and \reef{bin2}, the upper (lower) sign
corresponds to the portion of the boundary with $\x4>0$ ($\x4<0$).
Using these expressions and the dimensionless variables
\eqref{DimensionlessVariables}, the worldsheet action is
\begin{eqnarray}\label{actcurved}
I_\mt{WS} & = &{R^2\over \ell_s^2} \sqrt{u_0\over R} {1\over \pi}
\int_1^{z_\infty} \nts dz \, \psi(z,\sygn)+{3\over
8\pi}\int_1^{z_\infty} \nts dz \,\psi(z,\sygn)
\left(-{3\over 4 z^{1/2}}+{15 \sygn^3\over 4 z^{7/2}}\right)\log z
\nonumber \\
 &&+ {9\over 16 \pi}\psi_\infty(\sygn) \,z_\infty^{1/2} \left(1+{\sygn^3\over z_\infty^3}
\right)\,\log z_\infty + {39\over 16\pi}\int_1^{z_\infty} \nts dz \,
{\sqrt{1-\sygn^3}\over z\,\sqrt{z^8-\sygn^3 z^5-1+\sygn^3}}\log z
\nonumber \\
 &&+ {3\over 8}\,\log z_\infty + \log \left(g_s \left({u_0\over
 R}\right)^{3/4}\right)\,. \label{long}
 \end{eqnarray}
This expression must be renormalized according to the prescription
in section \ref{sec:renorm}, which in terms of the dimensionless
variables becomes
 \be
I_\mt{WS}^\mt{(ren)} = I_\mt{WS} -
\frac{L\,u_0}{2\,\pi\,\ap}\,z_{\infty} - \frac{9\,L\,\sqrt{u_{0}}}{16
\pi R^{3/2}}\,\sqrt{z_{\infty}} - \frac{3}{8}\, \log
 \left(z_{\infty}/z_\mt{sub}\right) \,.
 \label{normal}\ee
As described above, this prescription requires choosing a
subtraction scale $u_\mt{sub}=u_0\,z_\mt{sub}$. For simplicity, we
choose $u_\mt{sub} = u_0$ (\ie $z_\mt{sub}=1$) in the following. To
proceed, we must use \reef{noname} to simplify various factors, \eg
 \be
\frac{L\,u_0}{2\,\pi\,\ap}=\fc{\lambda_5}{2\pi\,L}\,\fc{\psi_\infty^2(\sygn)}{\pi}
 \ ,\qquad
\fc{R^2}{\ls^2}\sqrt{\fc{u_0}{R}}=\fc{\lambda_5}{2\pi\,L}\,\psi_\infty(\sygn)
\ .\label{shortening}
 \ee
In particular, with these expressions, the final result is expressed
as a function of the ratio $\lambda_5/L$ and the parameter $\sygn$. Then using \reef{normal}
to explicitly remove the divergent terms from \reef{actcurved}, the renormalized action becomes
 \be\label{FinalRenormAction}
I_\mt{WS}^\mt{(ren)} =
-\fc{\lambda_5}{2\pi\,L}\,\psi_\infty(\sygn)\, F_1(\sygn) - F_2
(\sygn) - \log \left(\pi N_c \right) + \frac{3}{2}\,\log
\left(\frac{\psi_\infty(\sygn)}{2\pi}\,\frac{\lambda_5}{L}\right)\,.
 \ee
where
 \ba\label{functions}
F_1(\sygn)&=& {1\over \pi} \int_1^{z_\infty} \nts \nts dz \,
\left(\psi_\infty(\sygn)-\psi(z,\sygn)\raisebox{11pt}{}\right)+{1\over \pi}\,\psi_{\infty}(\sygn)
 \\ \label{functionF2}
F_2(\sygn)&=& - \frac{9}{32\,\pi}\,\int_{1}^{z_{\infty}} \nts \nts
dz \,
\left(\psi_\infty(\sygn)-\psi(z,\sygn)\raisebox{11pt}{}\right)\,\frac{\log
z}{\sqrt{z}} + \frac{9}{8\,\pi}\,\psi_{\infty}(\sygn) \\ \nonumber &
&  - \frac{9\,\sygn^{3}}{16\,\pi}\,\psi_{\infty}(\sygn)\,\frac{\log
z_{\infty}}{z_{\infty}^{5/2}} -
\frac{45\,\sygn^{3}}{32\,\pi}\,\int_{1}^{z_{\infty}}\nts\nts dz\,
\psi(z,\sygn)\,\frac{\log z}{z^{7/2}} \\ \nonumber & & -
\frac{39}{16\,\pi}\,\int_{1}^{z_{\infty}}\nts\nts dz \,
\frac{\sqrt{1-\sygn^3}}{z\,\sqrt{z^{8}-z^{5}\,\sygn^{3}-1+\sygn^{3}}}\,\log
z ~.
 \nonumber
 \ea
The function $F_1(\sygn)$ is strictly positive, while $F_2(\sygn)$
is bounded from below. Further we note that the renormalization
indicated in \eqref{normal} has been incorporated in the definitions
of these functions in such a way each of the individual integrals
appearing in \eqref{functions} and \eqref{functionF2} is manifestly
finite.

The functions $F_1(\sygn)$ and $F_2(\sygn)$ appearing in the action
\eqref{FinalRenormAction} are obtained in general by numerically
performing the integrals in \eqref{functions} and
\eqref{functionF2}. However, in the extremal $D4$-brane background
(with $\ut=0$) an analytical expression can be given for the
renormalized worldsheet action \eqref{FinalRenormAction}. Using the
embedding \eqref{SUSYEmbedding} of the \branes in the extremal
background, the renormalized worldsheet action is
 \be \label{ExtremalAction}
I_\mt{WS}^\mt{(ren)}(\sygn=0) =
-\frac{1}{8\,\pi}\,\tan\left(\frac{\pi}{16}\right)\cdot\frac{\lambda_{5}}{L}
- \frac{3}{2}\,\log \left(\frac{L}{\lambda_5}\right) - \log N_c -
\frac{9}{64} +
\frac{3}{2}\,\log\left(\frac{2^{1/16}\,\Gamma(9/16)}{\pi^{7/6}\,\Gamma(1/16)}
\right) ~,
 \ee
in the limit that $z_\infty\ra\infty$. Thus, the expectation value of $\langle
\mathcal{O} \rangle$ takes the form
\begin{equation}
\langle {\cal O}\rangle \sim N_c \,
\left(\frac{L}{\lambda_5}\right)^{3/2} \, \exp \left
(\frac{1}{8\,\pi}\,\tan\left(\frac{\pi}{16}\right)\,\frac{\lambda_{5}}{L}
\right) \, . \label{oldowl}
\end{equation}
The exponential dependence on $\lambda_5 / L$ is precisely that
found in \cite{owl} coming from the Polyakov action. As described
above, the overall factor of $N_c$ comes from the Fradkin-Tseytlin
contribution \reef{actionx} to the action. This term also produces
the pre-factor of $(L/\lambda_5)^{3/2}$. However, one should keep in
mind that a complete calculation at this order in $\alpha'$
expansion would require evaluating the fluctuation determinant on
the string worldsheet. Hence one should expect this pre-factor to be
modified in a complete evaluation at this order.

Given the general result for the renormalized action
\reef{FinalRenormAction}, the expectation value of the operator
$\mathcal{O}$ is given by
  \be \label{O_g2N}
\langle {\cal O}\rangle \sim N_c \,
\left(\frac{L}{\lambda_5}\right)^{3/2} \,
\psi_{\infty}(\sygn)^{-3/2}\,\exp \left (
\frac{\lambda_5}{L}\,\frac{\psi_{\infty}(\sygn)}{2 \pi} \,F_1(\sygn)
+ F_2(\sygn)\right)\,.
  \ee
where implicitly we have again taken the limit $z_\infty\ra\infty$.
The functions $F_1(\sygn)$ and $F_2(\sygn)$ must be determined
numerically in the nonextremal background with $\ut\ne0$. As a check
of our numerical calculations, the action for the $\sygn=0$ case was
determined numerically and compared with \eqref{ExtremalAction}. In
addition, the Euler number was calculated numerically for each
$\sygn \neq 0$ embedding and compared with the expected value: $\chi
= 1$. In both cases the numerical error, expressed as a fraction of
the expected result, was of order of $10^{-10}$ to $10^{-11}$.

\FIGURE[ht]{
\includegraphics[width=.75\textwidth]{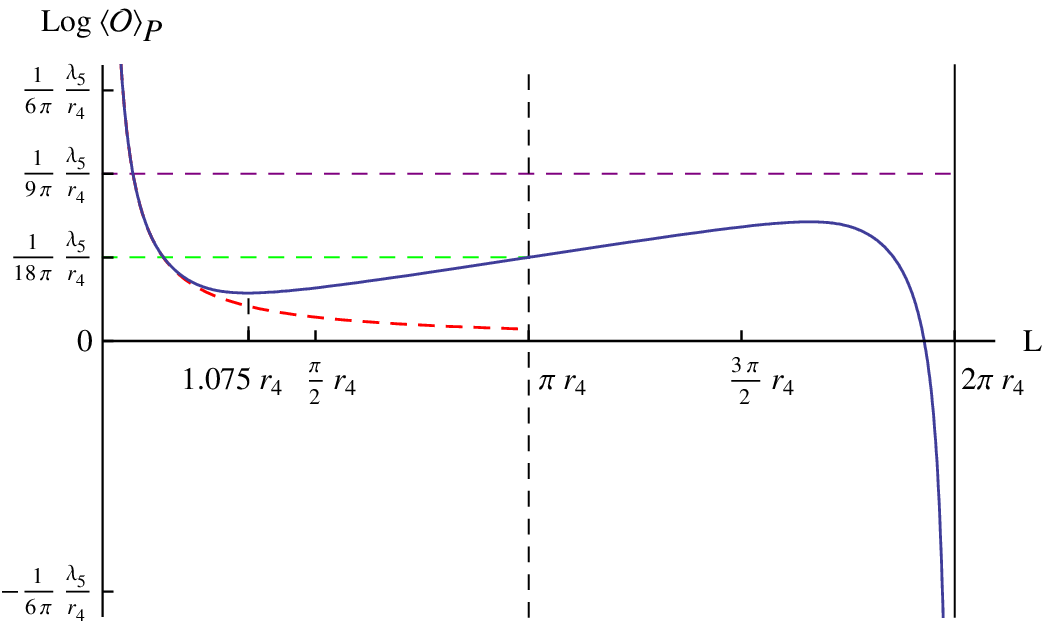}
\caption{The contribution to $\log \langle \mathcal{O} \rangle$ from
the renormalized Polyakov action as a function of $L$, with $r_4$
and $\lambda_5$ fixed. The dashed red curve is the result for the
extremal background. The dashed vertical line at $L=\pi\r4$
corresponds to the defects positioned at antipodal points on the
$\x4$ circle. The value of $\log \langle\mathcal{O}\rangle$ for this
configuration is indicated by the dashed green line. The dashed
purple line shows the value for the closed Wilson line $\langle
\mathcal{W}\rangle$.} \label{PvsL}}
As before, the Polyakov action (\ie the $F_1$ term in \reef{O_g2N})
dominates in the supergravity limit and so we focus on this term in
the following. The result depends on all three of the independent
parameters, $\lambda_5$, $\r4$ and $L$ (or rather dimensionless
ratios of these parameters) --- recall that $\sygn$ is implicitly
defined as a function of the ratio $\frac{L}{\r4}$ by the relation
\eqref{L_vs_uKK_relation}. A natural approach is to hold the gauge
theory parameters constant (by fixing the ratio $\lambda_5/\r4$) and
consider the expectation value as a function of $L$, the separation
of the defects. This is illustrated in figure \ref{PvsL}, where we
show $\log\langle \mathcal{O} \rangle\sim
\frac{\lambda_5}{L}\,\frac{\psi_{\infty}(\sygn)}{2 \pi}
\,F_1(\sygn)$ as a function of $L$. The plot shows that as $L/r_4$
approaches zero, our result follows the extremal result,
$\tan\left(\pi/16\right) \, \frac{\lambda_{5}}{8\pi\,L}$, appearing
in the exponential factor in \reef{oldowl}. This behaviour arises
because for $L \ll  \r4$, the $D8/\overline{D8}$ branes do not
extend very far into the bulk and so the string stretched between
them detects no difference between the extremal and non-extremal
backgrounds. In terms of the gauge theory, this behaviour simply
reflects the fact that the chiral symmetry breaking scale $U_0$ is
much larger than the confining scale $U_\mt{KK}$, where
supersymmetry breaking takes effect in the gauge theory. As $L$
becomes larger, the calculation begins to probe regions of the dual
spacetime geometry closer to $u=\ut$ and one sees that the extremal
and nonextremal behaviours of $\log\langle \mathcal{O} \rangle$
begin to deviate around $L\simeq\r4$. The expectation value reaches
an interesting local minimum at $L \sim 1.075 \, \r4$, where
$U_0/U_\mt{KK}\simeq1.608$. Note that the location of the minimum is
independent of $\lambda_5/\r4$ and is therefore always visible in
the supergravity limit.

When $L$ reaches $\pi\,\r4$, the defects are located at antipodal
points on the $\x4$ circle. This corresponds, in the relation
\eqref{L_vs_uKK_relation}, to the limiting value $\sygn=1$. However,
we have extended $L$ to the region $\pi\r4\le L\le 2\pi\r4$ in
figure \ref{PvsL}. Of course, in this regime, the shortest distance
between the defects on the $\x4$ circle is ${\widetilde
L}=2\pi\r4-L$ but the open Wilson line stretches the longer distance
$L$ around the circle.\footnote{In principle, the following
construction could be extended to consider Wilson lines which
connect the defects after fully winding around the $\x4$ circle some
number of times.} The embedding profile of the \branes is identical
to those in section \ref{pogo} but with $\widetilde L$ replacing
$L$. Now in the expectation value, the dual worldsheet spans the
minimal surface `outside' of the U-shape formed by the
D8-$\overline{\textrm{D8} }$ pair. Hence the (renormalized) Polyakov
action may be calculated as the action of a worldsheet covering the
entire $u$-$\x4$ geometry \reef{WSmet_and_dil} minus that for the
worldsheet stretched `inside' of the U-shape. As a result, at this
order, we have the relation: $\langle \mathcal{O}
\rangle({\widetilde L}/\r4=2\pi-L/\r4) \times \langle \mathcal{O}
\rangle(L/\r4)= \langle \mathcal{W}\rangle$ where $\langle
\mathcal{W}\rangle$ is the expectation value of a closed Wilson line
which winds once around the $\x4$ circle. (Note that we find
$\log\langle \mathcal{W}\rangle\simeq\lambda_5/(9\pi\r4)$.) Figure
\ref{PvsL} displays a symmetry about $L = \pi\,r_4$ which reflects
this relation and we may infer that $\langle \mathcal{O}
\rangle(L/\r4)$ approaches zero as $L\rightarrow2\pi\r4$. Of course,
we should add that the five-dimensional gauge theory is defined with
a cut-off $U_\infty$ and so one should not really consider the above
results for $L,{\widetilde L}\lsim 1/U_\infty$.

An alternative approach to considering the expectation value
$\langle \mathcal{O} \rangle$, is to consider it as a function of
$\r4$ with fixed $L$ and $\lambda_5$. The result is plotted in
figure \ref{Pvsr} for $\r4>L/\p$. In the decompactification limit,
$r_4 \to \infty$,  the expectation value again asymptotes to the
extremal result $\tan\left(\pi/16\right) \,
\frac{\lambda_{5}}{8\pi\,L}$. Note that, because $\sygn$ is
independent of the ratio $\lambda_5/L$, this plot can also be
understood as showing the dependence of $\langle \mathcal{O}
\rangle$ on the four-dimensional `t Hooft coupling, using the
relation $\lambda_4 = \lambda_5/(2\pi r_4)$.

%
\FIGURE[ht]{
\includegraphics[width=.75\textwidth]{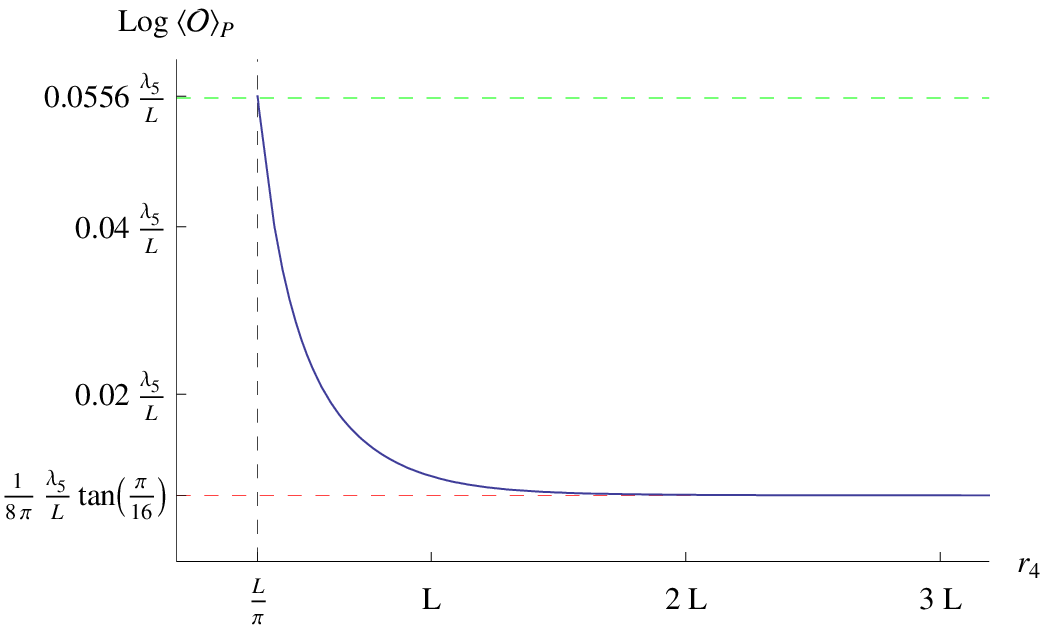}
\caption{The contribution to $\log \langle \mathcal{O} \rangle$ from
the  Polyakov action as a function of $r_4$, with $L$ and
$\lambda_5$ fixed. For large $r_4$ the value approaches the result
for the extremal background, shown as a dashed red line. }
\label{Pvsr}}

\section{Discussion}\label{discuss}

We have examined various aspects of a recent proposal
\cite{owl,mass1,old} to add quark masses to the Sakai-Sugimoto model
with nonlocal operators of the form \reef{oops}. The underlying
microscopic field theory is a five-dimensional gauge theory where
the chiral quarks are localized on separate four-dimensional
defects. However, the five-dimensional gauge theory is only defined
with a cut-off, \ie new degrees of freedom appear in the far UV. In
the dual supergravity background, this issue is realized by the
running of the dilaton which produces large string coupling in the
asymptotic region. In section \ref{wilson}, we examined
modifications introduced by the coupling of the dilaton to the
string worldsheet. In particular, we showed that this coupling calls
for a modification of the renormalization of these operators as in
\reef{fullir}. The first two subtractions, which are linear in the
length $L$, renormalize the Wilson line and are not particular to
the present open Wilson line calculations. Hence both of these
terms, including the second one proportional to $\sqrt{\uin}$, would
appear in calculations for closed Wilson lines as well. On the other
hand, the log subtraction is distinctive of the two end-points of
the open Wilson line.

It is interesting to re-express the subtractions in \reef{fullir} in
terms of an energy cut-off, using the standard dictionary $U_\infty
=\uin/2\pi\ls^2$,
 \be
 I_\mt{WS}^\mt{(ren)} = I_\mt{WS} - L\,U_{\infty} -
 \frac{9}{4\sqrt{2}}\,\frac{L\,U_{\infty}}{g_{eff}(U_{\infty})}
-\frac{3}{8}\, \log \left(\fc{U_{\infty}}{U_\mt{sub}}\right) \,.
 \label{fullirx}
 \ee
In the third term, $g_{eff}(U_{\infty})$ is the (dimensionless)
effective coupling \reef{run} of the five-dimensional gauge theory
evaluated at the cut-off scale $U_{\infty}$. Hence the $\alpha'$
expansion on the string worldsheet produces an expansion in inverse
powers of the coupling $g_{eff}$ from the gauge theory perspective,
rather than the $1/\nc$ expansion as produced by
$\alpha'$-corrections to the supergravity action --- a similar
observation was made about the thermal quark diffusion constant in
\cite{spectral}. It is interesting that the energy scale
$U_{\infty}/g_{eff}(U_{\infty})$ appearing in the second subtraction
is the supergravity energy scale associated with $\uin$ \cite{Peet}.
This is a natural energy scale to appear here since fluctuations on
the worldsheet are contributing at this order \cite{holom}.

In section \ref{sec:SmoothWS}, we explicitly calculated the
expectation value of the nonlocal fermion bilinear. This expectation
value characterizes the chiral condensate in this holographic model.
As this holographic construction does not permit the construction of
a local fermion bilinear, this expectation value is the best order
parameter to characterize the spontaneous breaking of chiral
symmetry. Our explicit calculations yield the result given in
\reef{O_g2N}. We note that \reef{R} and \reef{noname} can be used to
express the pre-factor of the Polyakov term as $\lambda_5/L\sim
g_{eff}(U_0)$, up to numerical factors, where $g_{eff}(U_0)$ is the
effective coupling evaluated at the chiral symmetry breaking scale
$U_0$. The dependence of $\langle {\cal O}\rangle$ on $L/\r4$
illustrated in figure \ref{PvsL} describes the intricate interplay
of the supersymmetry breaking (or confinement) and chiral symmetry
breaking scales in determining the expectation value. Of course, in
the absence of supersymmetry breaking, the result in the extremal
background \reef{oldowl} is independent of $L/\r4$ \cite{owl}. As
figure \ref{PvsL} also illustrates, $\langle{\cal O}\rangle$
approaches this supersymmetric result in the limit $L/r_4 \ra 0$.

Studying the theory at finite temperature in this regime, one finds
that the chiral symmetry breaking and confinement/deconfinement
phase transitions are independent \cite{aharony,parnachev}. As
described in section \ref{sec:Background}, the chiral symmetry
breaking is realized in the gravitational dual by the \branes
joining together to form a smooth U-shaped embedding. The deconfined
phase of the gauge theory is represented by replacing the
supergravity background by a D4 black hole \cite{thermalwitten}. The
transition between the low-temperature confining phase and the
high-temperature deconfined phase occurs when \cite{thermalwitten}
 \be
\ct=\fc{1}{2\pi\,\r4}\,.\label{transit1}
 \ee
In the deconfined phase, if $u_0$ is sufficiently large, the tension
of the D8-branes can support the U-shaped embedding against
gravitational attraction of the black hole, which has the
interpretation that the chiral symmetry remains broken in the
deconfined phase \cite{aharony,parnachev}. Chiral symmetry is not
restored until a temperature given by
 \be T_{\chi\mt{SB}} = \frac{1}{2\pi
r_4}\,\frac{L_c}{L} \simeq \frac{0.154}{L}
 \ee
where $L_c=0.97\r4$. Above this temperature, the gravitational
attraction becomes sufficiently large that the \branes are pulled
into the horizon (and the embedding is trivial, \ie $\x4=constant$),
as shown in figure \ref{restoredsymmetry}. The phase structure of
the Sakai-Sugimoto model is summarized in figure \ref{phase}.
\FIGURE[ht]{
\includegraphics[width=.75\textwidth]{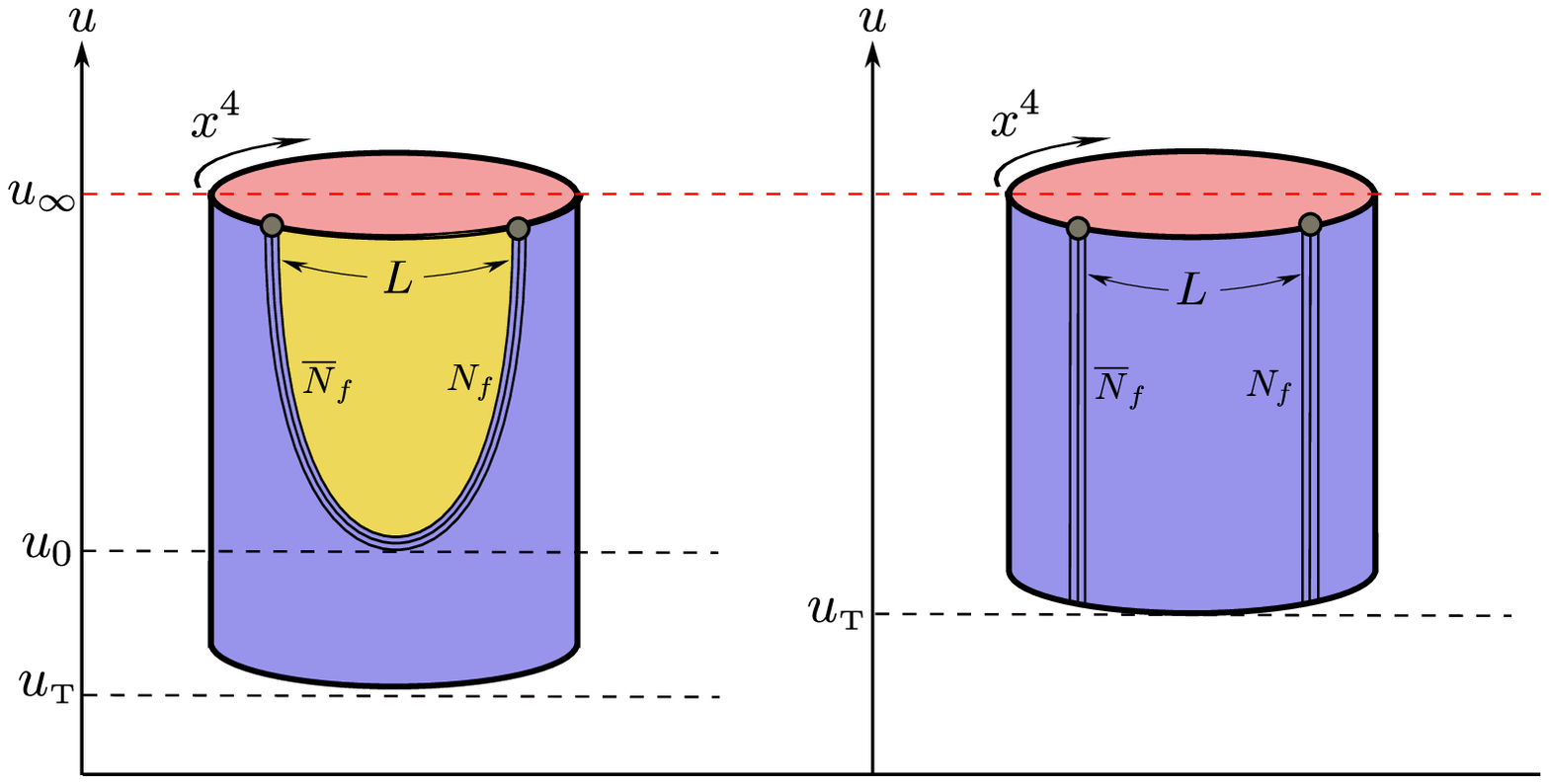}
\caption{For small $L$, $u_0$ is sufficiently large that the \branes
remains outside of the black hole horizon at $\utt$, as shown on the
left. For larger $L$, the branes fall through the horizon, as shown
on the right.} \label{restoredsymmetry}}

In the high temperature phase, where the \branes is disconnected,
the chiral symmetry is restored and so this should be reflected in
the expectation value. In particular, beyond the phase
transition of \cite{aharony,parnachev}, one should have $\langle \OO
\rangle=0$. In fact, this result does arise because with the trivial
embedding in the black hole background there is no string worldsheet
connecting the \branes with a single asymptotic boundary. The
simplest consistent worldsheet would extend through the
`Einstein-Rosen' throat and out to the boundary of the second
asymptotic region in the black hole geometry. Hence this worldsheet
would be relevant for a correlator of two operators with the second
being in the thermofield double of the original gauge theory
\cite{eternal}.
\FIGURE[ht]{
\includegraphics[width=.75\textwidth]{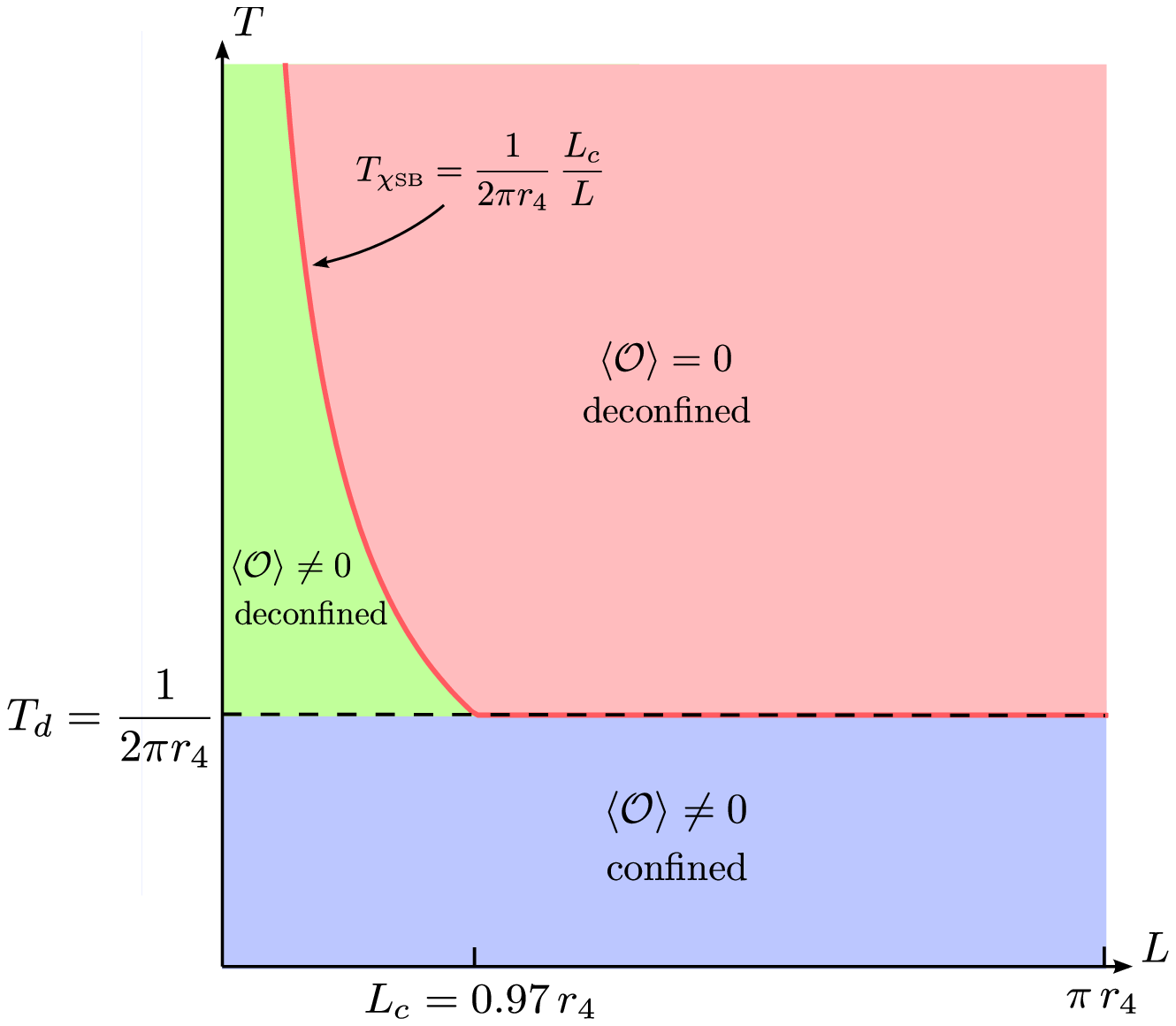}
\caption{The phase diagram for the Sakai-Sugimoto model. The transition from confinement to deconfinement occurs at $T_d = (2\pi r_4)^{-1}$. Chiral symmetry is also restored at this temperature if $L > 0.97\,r_4$. The transitions are independent for $L < 0.97 \,r_4$, in which case chiral symmetry is restored above $T_{\chi\mt{SB}} = .154/L$.} \label{phase}}

In a similar way, these expectation values are useful for
characterizing the different phases in theories with many defects,
as discussed, \eg in \cite{manyd}. Again, one would find $\langle
\OO \rangle=0$ for a Wilson line operator connecting two defects
which are not dual to a \branes which are not joined.

As observed at the end of section \ref{sec:RectWS}, $\langle \OO
\rangle\propto \nc$ in accord with the standard large $\nc$
counting. In our calculations, we essentially set the number of
flavours to one, however, if $\nf>1$ one might anticipate the
expectation values would also be proportional to $\nf$, again
reflecting the number of degrees of freedom involved in such a
bilinear --- \eg, see \cite{brain,findens}. However, in the case
where $\nf>1$, we are implicitly considering the expectation value
$\langle \OO^{IJ}\rangle$ of an operator with flavour indices $I$
and $J$ for the two fermions. For the smooth embeddings, we would
have $\langle \OO^{IJ}\rangle\propto \delta^{IJ}$ because
consistency requires that the worldsheet start and end on the same
brane throughout the embedding. A priori, there is no connection
between the $\psi_L^I$ on one defect and the $\psi_R^J$ on the
other. So our operator reveals this connection as established by the
chiral symmetry breaking. Tracing over the flavour indices would
correspond to implicitly summing over the different worldsheets and
would produce the factor of $\nf$ mentioned above.

An alternate approach to understanding chiral symmetry breaking in
the Sakai-Sugimoto model was considered in
\cite{tach1,tach2,tach3,tach4}. There the key element is the open
string tachyon that develops between the \branes when the (proper)
distance separating them is small. Chiral symmetry breaking is
realized as the condensation of the tachyon, which leads to
brane-anti-brane annihilation deep in the IR region, producing the
smooth embedding in which the \branes join. The quark mass and the
chiral condensate would be related to the asymptotically growing and
decaying modes of the tachyon field. This description and the
approach examined in the present paper both consider the physics of
open strings stretched between the D8- and
$\overline{\textrm{D8}}$-branes, so it seems that they must be
related. Conceptually, one can think of the tachyon analysis as the
second-quantized description of the relevant open string physics
while the worldsheet procedure \cite{owl,mass1,old} considered above
is the first-quantized description of essentially the same physics.
Of course, it would be interesting to make this connection more
precise. This naturally calls for a proper quantization of (open)
strings in the supergravity background of the D4-brane throat. A
more accessible route may be to examine the D8-brane embeddings for
a nonvanishing quark mass, following the suggestion of \cite{owl} to
include the Polyakov action for the instantonic worldsheet as part
of the action for the \branes. One could then consider the
dependence of $\langle\OO\rangle$ on $\mq$ and compare with the
results given in \cite{tach2,tach3,tach4}.


\acknowledgments It is a pleasure to thank Ofer Aharony, Martin
Kruczenski, David Kutasov and Arkady Tseytlin for useful
correspondence and conversations. RCM would also like to thank
Shigeki Sugimoto and especially Rowan Thomson for their
collaboration at a very early stage of this project. Research at
Perimeter Institute is supported by the Government of Canada through
Industry Canada and by the Province of Ontario through the Ministry
of Research \& Innovation. RCM also acknowledges support from an
NSERC Discovery grant and funding from the Canadian Institute for
Advanced Research.

\appendix

\section{Fluctuation Determinant} \label{fluctuate}

In this appendix we will evaluate the contribution from fluctuations
for rectangular worldsheet discussed in section \ref{sec:RectWS}.
Since $\alpha'$ is a loop-counting parameter on the string
worldsheet, the one-loop fluctuation determinant will contribute at
the same order as the Fradkin-Tseytlin term \reef{actionx},
considered in the main text. The primary result here is to show that
the fluctuation determinant contributes no additional UV divergences
to the Wilson line calculations, at this order. Since the UV
behaviour is universal, this result applies for all of the Wilson
line calculations considered in the present paper. A similar
analysis of worldsheet fluctuations for a standard closed Wilson
line in the D4-brane background has been performed in \cite{leo}. To
begin the calculation, we must expand the worldsheet action to
quadratic order about a classical solution, $X^i$. Working with the
Green-Schwarz formalism,\footnote{One might question whether or not
the Fradkin-Tseytlin term \reef{actionx} is to be added in the
worldsheet action of the Green-Schwarz string. While the classical
action does not couple to the dilaton, this interaction is still
necessary at the quantum level  to preserve the conformal and
$\kappa$ symmetry of Green-Schwarz string, just as in the bosonic
case (and also to have proper effective string coupling dependence
of string loops).} this yields
 \be I=I_{B}+I_F+I_{ghosts}\,,
 \ee
where
 \ba
I_B={1\over 4\pi}\int d^2 \sigma \sqrt{g} (G_{ij}D_\alpha \xi^i
D_\beta \xi^j g^{\alpha \beta}&+&R_{ik_1 k_2 j}\xi^{k_1} \xi^{k_2}
\partial_\alpha X^i \partial_\beta X^j
g^{\alpha\beta}\nonumber\\ &+& R^{(2)}[\ell_s D_i\Phi \xi^i+{1\over
2} \ell_s^2 D_i D_j \Phi \xi^i \xi^j])\,,\label{crusty}
 \ea
 \be I_F={i\over 4\pi}\int d^2\sigma \bar
\Theta[(\sqrt{-h}h^{\alpha\beta}+\epsilon^{\alpha\beta}\Gamma^{11})\Gamma_\alpha
{\cal D}_\beta]\Theta\,,
 \ee
and
 \be I_{ghosts}={1\over 2}\int d^2\sigma
\sqrt{g}g^{\alpha\beta}(g^{\gamma\delta}\nabla_\gamma
\epsilon_\alpha\nabla_\delta\epsilon_\beta-{1\over
2}R^{(2)}\epsilon_\alpha \epsilon_\beta)\,.
 \ee
In the above, $i$ labels a spacetime index which we will split in
what follows as $a$ labeling a $S^4$ direction and $\mu$ labeling
the remaining transverse directions. The worldsheet directions will
be labeled by lower Greek indices. The other quantities are
specified by\cite{leo}
 \be
D_\alpha \xi^j=
\partial_\alpha X^i (\partial_i\xi^j+\Gamma^j_{il}\xi^l)\,,
 \ee
 \be
{\cal D}_i=\partial_i+{1\over 4}\omega_{i a b}\Gamma^{ab}-{1\over
8\cdot 4!}e^\phi F_{abcd}\Gamma^{abcd}\Gamma_i\equiv
\partial_i+M_i\,,
 \ee
where $a,b,c,d$ are tangent space indices. The $\kappa$-symmetry
transformation for the GS fermions is given by
 \be
\delta_\kappa\Theta=(1-{\epsilon^{\alpha\beta}\over
2\sqrt{-g}}\Gamma_{\alpha\beta}\Gamma^{11})\kappa\,.
 \ee
Note that for simplicity in the following calculations, we will set
$R=1$ in the supergravity background. We will also drop the dilaton terms in what follows since they are subleading.

\noindent{\bf {Fermion contributions}}

We will use the zehnbeins
 \ba
e^\mu&=&u^{3/4} dx^\mu\,,\quad \mu=1..4\\
e^0 &=& u^{3/4} dt \,, \quad e^5= u^{-3/4} d u\,,\\
e^6 &=& u^{1/4} d\psi\,, \quad e^7=u^{1/4} \cos\psi d\chi\,,\\
e^8 &=& u^{1/4} \cos\psi\sin\chi d\phi_1\,, \quad
e^9=u^{1/4}\cos\psi\cos\chi d\phi_2\,.
 \ea
In the following we will sometimes use the notation
$\mu_6=1,\mu_7=\cos\psi,\mu_8=\cos\psi\sin\chi,
\mu^9=\cos\psi\cos\chi$. The RR field strength in tangent space is
given by
 \be
F_{6789}={3\over u}\,.
 \ee
In the case at hand, $x^4=\sigma_0$ and $u=\sigma_1$ where
$\sigma_0,\sigma_1$ are worldsheet variables. Using the pullback
metric for the Euclidean worldsheet and noting that
$\sqrt{-g}=i\sqrt{|g|}$ we get
 \be
I_F={i\over 4\pi}\int d^2\sigma \left(\bar\Theta
\Gamma^{11}(\Gamma_u{\cal D}_0-\Gamma_{x^4}{\cal D}_1)\Theta-i
\bar\Theta (\sigma_1^{3/2}\Gamma_u{\cal
D}_1-\sigma^{-3/2}\Gamma_{x^4}{\cal D}_0)\Theta\right)]\,.
 \ee
Here we have absorbed $\alpha'$ into the fluctuations. Using
 \ba
\Gamma_u M_0-\Gamma_\tau M_1 &=& -{3\over 8} u^{-1/4}\Gamma_4
+{3\over 4}u^{-1/4}\tilde\Gamma\Gamma_{45}\,,\\
\Gamma_u M_1 &=& -{3\over 8}u^{1/4} \tilde \Gamma\,,\\
\Gamma_\tau M_0&=& {3\over 8}u^{5/4}-{3\over 8} u^{5/4}\tilde
\Gamma\,,
 \ea
with $\tilde \Gamma=\Gamma^{6789}$ in tangent space we get
 \ba
I_F={i\over 4\pi}\int
d^2\sigma&\bigg{(}&u^{-3/4}\bar\Theta(\Gamma^{11}\Gamma_5-i
\Gamma_4)\partial_0\Theta+u^{3/4}\bar\Theta(-i\Gamma_5-\Gamma^{11}\Gamma_4)
\partial_1\Theta\nonumber\\&-&{3u^{-1/4}\over 8}
\bar\Theta(\Gamma^{11}\Gamma_4-i\Gamma_5)\Theta+{3
u^{-1/4}\over
4}\bar\Theta(i\tilde\Gamma+\Gamma^{11}\tilde\Gamma\Gamma_{45})\Theta\bigg{)}\,.
 \ea
We will fix $\kappa$-symmetry with the following: First split
$\Theta=\theta_1+\theta_2$. Then choose
 \be
\Gamma_5\theta^1=i\Gamma_4\theta^1\,,\quad
\Gamma_5\theta^2=-i\Gamma_4\theta^2\,,
 \ee
which leads to after redefining
$\hat\theta^{1,2}=u^{3/8}\theta^{1,2}$ and defining
$\partial_\pm=\mp i\sigma^{-3/2}\partial_0+\partial_1$
 \be
I_F={i\over 2\pi}\int d^2\sigma\left(\bar{\hat\theta}^1\Gamma_4
\partial_+\hat\theta^1-\bar{\hat\theta}^2\Gamma_4\partial_-\hat\theta^2+{3i\over
4\sigma_1} (\bar{\hat\theta}^1\tilde\Gamma
\hat\theta^2+\bar{\hat\theta}^2\tilde\Gamma\hat\theta^1)\right)\,.
 \ee
Now choosing the gamma matrices $\Gamma_a$ such that
 \be
\Gamma_0=\tau_2\otimes 1\,, \Gamma_1=\tau_1\otimes
1,\Gamma_A=\tau_3\otimes\gamma_A
 \ee
with $\gamma_A$ being Euclidean Dirac matrices in 8 dimensions and
splitting the $\theta^I$'s into two Euclidean Majorana-Weyl
fermions of opposite chiralities $S,\tilde S$ we get
 \be
I_F={i\over 2\pi}\int d^2\sigma \left({S}\partial_+ S-{\tilde
S}\partial_-\tilde S+{3\over 4\sigma_1}(S\tilde\gamma\tilde
S-{\tilde S}\tilde \gamma S)\right)\,.
 \ee
The squared equations of motion following from the above are:
 \ba
(\partial_0^2+\sigma_1^{3/2}\partial_1\sigma_1^{3/2} \partial_1
-{\sqrt{\sigma_1}\over 2}(i\partial_0+\sigma^{3/2}\partial_1)-{9\over 16}{\sigma_1})S&=&0\,,\\
(\partial_0^2+\sigma_1^{3/2}\partial_1\sigma_1^{3/2}\partial_1-{\sqrt{\sigma_1}\over
2}(-i\partial_0+\sigma^{3/2}\partial_1)-{9\over 16}{\sigma_1})\tilde
S&=&0\,.
 \ea
Combining $S,\tilde S$ into a worldsheet spinor $\Psi=(S,\tilde S)$,
the equations of motion for the fermions then takes on the form
 \be
(\partial_0^2+\sigma_1^{3/2}\partial_1\sigma_1^{3/2}\partial_1-{9
\sigma_1\over 16}+{3\sigma_1\over 8}\tau^1\tilde\gamma)\Psi=0\,.
 \ee
Here $\tau^1$ represents the Pauli matrix $((0,1),(1,0))$. In first
order perturbation theory, the term proportional to $\tau_1$ will
not contribute. Since $S$ and $\tilde S$ combine to form a
worldsheet spinor, we have 8 massive fermions satisfying the above
equation\footnote{Otherwise naively it would appear that there are 8
$S$'s and 8 $\tilde S$'s giving 16 fermions which would lead to the
wrong counting.}.

\noindent{\bf Boson contributions}

In the second line of \reef{crusty}, we have included the
contributions of the Fradkin-Tseytlin term \reef{actionx}. However,
these two terms come with explicit factors of the string length
$\ls$ which reflects the fact that they would only contribute in a
two-loop calculation of the fluctuation determinant. Therefore we
ignore these last two contributions in the following calculation.
The quadratic order action for the bosons is given by
 \ba
I_B={1\over 4\pi}\int d^2\sigma{1\over  \sigma_1^{3/2}}&
\bigg{(}&\dot{\hat\xi}_u^2+\sigma_1^3\hat\xi_u'^2+{15\sigma_1\over 16}
\hat\xi_u^2+\dot{\hat\xi}_{x^4}^2+\sigma_1^3\hat\xi_{x^4}'^2
+{15\sigma_1\over 16}\hat\xi_{x^4}^2-24\sqrt\sigma_1 (\hat\xi_{x^4}
\dot{\hat\xi}_u-\hat\xi_u\dot{\hat\xi}_{x^4})\nonumber \\
&+&\dot{\hat\xi}_\mu^2+\sigma_1^3\hat\xi_\mu'^2+{15\sigma_1\over
16}\hat\xi_\mu^2+\mu_a^2(\dot{\hat\xi}_a^2+\sigma_1^3\hat\xi_a'^2+{3\sigma_1\over
16}\hat\xi_a^2)\bigg{)}\,.
 \ea
Here we have defined $\hat\xi_{x^4}=u^{3/4}\xi_{x^4},
\hat\xi_u=u^{-3/4}\xi_u,\hat\xi_\mu=u^{3/4}\xi_\mu,\hat\xi_a=u^{1/4}\xi_a$.
Thus the mass terms are at $O(1/\sqrt{\sigma_1})$. Hence we now have
$\hat\xi_u,\hat\xi_{x^4},\hat\xi_\mu$ satisfying
 \be
(\partial_0^2+\sigma_1^{3/2}\partial_1\sigma_1^{3/2}\partial_1-{15\sigma\over
16})\hat\xi=0\,, \ee while $\hat\xi_a$ satisfy \be
(\partial_0^2+\sigma_1^{3/2}\partial_1\sigma_1^{3/2}-{3\sigma\over
16})\hat\xi=0\,,
 \ee

\noindent{\bf Ghost contributions}

The ghost action works out to be
 \be
I_{ghosts}={1\over 2}\int d^2\sigma {1\over
\sigma_1^{9/2}}(\dot{\hat\epsilon}_1^2+\sigma_1^3\hat\epsilon_1'^2+{15\sigma\over
16}\hat\epsilon_1^2+\sigma_1^6(\dot{\hat\epsilon}_2^2+\sigma_1^3\hat\epsilon_2'^2+{15\sigma\over
16}\hat\epsilon_2^2)-24\sigma^{7/2}(\hat\epsilon_2\dot{\hat\epsilon}_2-
\hat\epsilon_1\dot{\hat\epsilon}_2))\,,
 \ee
with $\hat\epsilon_1=\sigma_1^{3/4}\epsilon_1$ and
$\hat\epsilon_2=\sigma_1^{-3/4}\epsilon_2$. The ghosts satisfy
 \ba
(\partial_0^2+\sigma_1^{9/2}\partial_1\sigma_1^{-3/2}\partial_1-
{15\sigma_1\over 16})\hat\epsilon_1 &=&0\,,\\
(\partial_0^2+\sigma_1^{-3/2}\partial_1\sigma_1^{9/2}\partial_1-{15\sigma_1\over
16})\hat\epsilon_2 &=&0\,.
 \ea

\noindent{\bf Final Result}

The result for the partition function after the above laborious
calculation is
 \be \label{det}
{\det^{8/2}(\partial_t^2+\sigma^{3/2}\partial_\sigma
\sigma^{3/2}\partial_\sigma-{9\over
16}\sigma)\det^{1/2}(\partial_t^2+\sigma^{9/2}\partial_\sigma
\sigma^{-3/2}\partial_\sigma-{15\over
16}\sigma)\det^{1/2}(\partial_t^2+\sigma^{-3/2}\partial_\sigma
\sigma^{9/2}\partial_\sigma-{15\over 16}\sigma)\over \det^{6/2}
(\partial_t^2+\sigma^{3/2}\partial_\sigma
\sigma^{3/2}\partial_\sigma-{15\over
16}\sigma)\det^{4/2}(\partial_t^2+\sigma^{3/2}\partial_\sigma
\sigma^{3/2}\partial_\sigma-{3\over 16}\sigma)}\,.
 \ee
The evaluation of the determinant exactly is in general a very hard
problem \cite{dunne}. In this case we note that (\ref{det}) can be
rewritten as
 \be
{\det^{8/2}(\partial_t^2+\partial_x^2-{f \over
x^2})\det^{1/2}(\partial_t^2+\partial_x^2-{g_1\over
x^2})\det^{1/2}(\partial_t^2+\partial_x^2-{g_2\over x^2})\over
\det^{6/2} (\partial_t^2+\partial_x^2-{b_1\over x^2})
\det^{4/2}(\partial_t^2+\partial_x^2-{b_2\over x^2})}\,,
 \ee
where $x\equiv-{2\over \sqrt{\sigma}}$ and $f=9/4, g_1=39/4,
g_2=63/4, b_1=15/4, b_2=3/4$. Then each of the operators featuring
in the determinant can be written as \cite{dunne}
 \be
D=\partial_x^2 - {(l-1/2)(l+1/2)\over x^2}\,,
 \ee
so that we can write the determinant as $\prod \lambda_n$ where
$\lambda_n$ is given by
 \be\label{eg}
(-D+{4 m^2\pi^2\over L^2}) \phi_n=-\lambda_n \phi_n\,.
 \ee
The function $\phi_n$ satisfies Dirichlet boundary conditions,
namely $\phi_n(x=0)=0, \phi_n(x=x_0)=0$. Here $x_0$ is related to $u_0$
through $x_0=-2/\sqrt{u_0}$. The solution to (\ref{eg}) with the
Dirichlet boundary conditions are known to be Bessel functions
$\sqrt{x} J_l (x)$, $\sqrt{x} Y_l(x)$. Since both $\pm l$ are
allowed, we will choose $\sqrt{x} J_l(x)$ and $\sqrt{x} Y_{-l}(x)$
to be the independent solutions. Then imposing the boundary
condition at $x=x_0$ we have
 \begin{eqnarray}
J_l (\omega_n x_0)&=&0\,, \qquad \omega_n^2=({j_{l,n}\over x_0})^2
=\lambda_n-{4 m^2\pi^2\over L^2}\,,\\
Y_{-l}(\omega_n x_0) &=& 0\,, \qquad \omega_n^2=({y_{-l,n}\over
L})^2=\lambda_n-{4 m^2\pi^2\over L^2}\,,
 \end{eqnarray}
so that $\omega_n$'s are related to the zeros $j_{l,n}, y_{-l,n}$ of
the Bessel functions. Then the determinant can be written using the formula
\be \sinh x=x \prod_{k=1}^\infty (1+{x^2\over
k^2 \pi^2})\,,
\ee
as
 \be \label{exact}
{\cal D}={\cal D}_j {\cal D}_y\,,
 \ee
where
 \begin{eqnarray}
{\cal D}_j&=&\prod{\sinh^4 \displaystyle {j_{l_f, n_f}L\over 2
|x_0|} \sinh^{1/2} {j_{l_{g_1}, n_{g_2}}L\over 2 |x_0|}\sinh^{1/2}
{j_{l_{g_2}, n_{g_2}}L\over 2 |x_0|}  \over \sinh^3 \displaystyle
{j_{l_{b_1}, n_{b_1}}L\over 2 |x_0|} \sinh^2 {j_{l_{b_2},
n_{b_2}}L\over 2 |x_0|}}\,.
 \\
{\cal D}_y&=&\prod{\sinh^4 \displaystyle {y_{-l_f, n_f}L\over 2
|x_0|} \sinh^{1/2} {y_{-l_{g_1}, n_{g_2}}L\over 2 |x_0|}\sinh^{1/2}
{y_{-l_{g_2}, n_{g_2}}L\over 2 |x_0|}  \over \sinh^3
\displaystyle {y_{-l_{b_1}, n_{b_1}}L\over 2 |x_0|} \sinh^2
{y_{-l_{b_2}, n_{b_2}}L\over 2 |x_0|}}\,.
 \end{eqnarray}
Now we want to get the large $n$ asymptotics of this function. We
use the useful identity\cite{watson} that the large zeros of the
Bessel function behave as
 \be
j_{\nu,n} \cos \alpha-y_{\nu,n}\sin\alpha \approx (n+{\nu\over
2}-{1\over 4})\pi-\alpha-{4\nu^2-1\over 8[(n+{\nu\over 2}-{1\over
4})\pi-\alpha]}+\cdots\,,
 \ee
to get
 \begin{eqnarray}
\log {\cal D}\approx 0.09 \pi \sum_n \bigg{(} 8(n\pm{1\over
2}{\sqrt{5\over 2}}\mp{1\over 4}) &+&(n\pm{1\over
2}\sqrt{10}\mp{1\over 4})+(n\pm2\mp{1\over 4})\nonumber \\&-&6(n\pm 1\mp {1\over
4})-4(n\pm {1\over 2}\mp{1\over 4})\bigg{)}\,,
 \end{eqnarray}
using which the leading divergence cancels. The subleading terms
arise from $O(1/n^2)$ terms which lead to a finite result at
$O(1/u_\infty^{9/2})$. The exact formula (\ref{exact}) allows us in
principle to extract this finite number although we will not attempt
it here, as this contribution would vanish in the relevant limit
$u_\infty\ra\infty$. Hence our key result is that in the fluctuation
determinant \reef{det} is in fact precisely 1 in this limit.

\end{document}